\documentclass[sigplan,review,10pt]{acmart}
\renewcommand\footnotetextcopyrightpermission[1]{}
\usepackage{xurl}
\usepackage{paralist}
\usepackage{graphicx}
\usepackage{color}
\usepackage{hyperref}
\usepackage{breakurl}
\usepackage{comment}
\usepackage{subfigure}
\usepackage{xspace}
\usepackage{pgfplots}
\usepackage{booktabs}

\usepackage[T1]{fontenc}
\usepackage{lmodern}

\pgfplotsset{compat=1.9}

\setcopyright{acmcopyright}

\newcommand{\confname}[3]{
\ifx\longconferencenames\undefined
\newcommand{#1}[0]{{#2}}
\else
\newcommand{#1}[0]{{#3}}
\fi
}

\newcommand{\cvnote}[1]{
\ifx\includecvnotes\undefined
\else
{#1}%
\fi
}

\newcommand{\bionote}[1]{
\ifx\includebionotes\undefined
\else
{#1}%
\fi
}

\confname{\podc}{PODC}{Proceedings of the ACM symposium on Principles of
distributed computing (PODC)}

\confname{\asplos}{ASPLOS}{{Proceedings of the ACM International Conference on Architectural Support for Programming Languages and Operating Systems (ASPLOS)}}

\confname{\spaa}{SPAA}{{Proceedings of the ACM symposium on Parallelism in algorithms and architectures (SPAA)}}

\confname{\osdi}{OSDI}{{Proceedings of the USENIX Symposium on Operating Systems Design and Implementation (OSDI)}}

\confname{\disc}{DISC}{{Proceedings of the International Conference on Distributed Computing (DISC)}}

\confname{\usenixatc}{USENIX}{{Proceedings of the USENIX Annual Technical Conference}}
\confname{\usenixsec}{USENIX Security}{{Proceedings of the USENIX Security Symposium}}

\confname{\pldi}{PLDI}{{Proceedings of the ACM SIGPLAN conference on Programming language design and implementation (PLDI)}}

\confname{\computer}{Computer}{{IEEE Computer}}

\confname{\sosp}{SOSP}{{Proceedings of the ACM SIGOPS Symposium on Operating Systems Principles (SOSP)}}

\confname{\isca}{ISCA}{{Proceedings of the ACM IEEE International Symposium on Computer Architecture (ISCA)}}

\confname{\csaw}{CSAW}{{Proceedings of the ACM Workshop on Computer Security Architecture (CSAW)}}

\confname{\wddd}{WDDD}{{Proceedings of the Workshop on Duplicating, Deconstructing, and Debunking (WDDD)}}

\confname{\vldb}{VLDB}{{Proceedings of the International Conference on Very Large Databases (VLDB)}}

\confname{\toplas}{TOPLAS}{{ACM Transactions on Programming Languages and Systems (TOPLAS)}}

\confname{\tocs}{TOCS}{{ACM Transactions on Computer Systems (TOCS)}}

\confname{\ppopp}{{PPoPP}}{{Proceedings of the ACM SIGPLAN Symposium on Principles and Practice of Parallel Programming (PPoPP)}}

\confname{\jpdc}{J. Parallel Distrib. Comput.}{{Journal of Parallel and Distributed Computing}}

\confname{\ismm}{ISMM}{{Proceedings of the ACM International Symposium on Memory Management (ISMM)}}

\confname{\cacm}{CACM}{{Communications of the ACM (CACM)}}

\confname{\hpca}{HPCA}{{Proceedings of the IEEE International Symposium on High-Performance Computer Architecture (HPCA)}}

\confname{\transact}{TRANSACT}{{Proceedings of the ACM SIGPLAN Workshop on Transactional Computing (TRANSACT)}}

\confname{\iiswc}{IISWC}{{Proceedings of the IEEE International Symposium on Workload Characterization (IISWC)}}

\confname{\tpds}{IEEE Trans, Parallel Distrib. Syst.}{{IEEE Transactions on Parallel and Distributed Systems}}

\confname{\osr}{OSR}{{ACM Operating Systems Review}}

\confname{\nsdi}{NSDI}{{Proceedings of the USENIX Symposium on Networked Systems Design and Implementation (NSDI)}}

\confname{\cc}{CC}{{Proceedings of the International Conference on Compiler Construction (CC)}}

\confname{\surveys}{ACM Comput. Surv.}{{ACM Computing Surveys}}
\confname{\icde}{IDCE}{{Proceedings of the IEEE International Conference on Data Engineering (ICDE)}}
\confname{\fast}{FAST}{{Proceedings of the USENIX Conference on File and Storage Technologies (FAST)}}
\confname{\eurosys}{{E}uro{S}ys}{{Proceedings of the ACM European Conference on Computer Systems ({E}uro{S}ys)}}
\confname{\hotos}{HotOS}{{Proceedings of the USENIX Workshop on Hot Topics in Operating Systems (HotOS)}}
\confname{\hotosnew}{HotOS}{{Proceedings of the ACM Workshop on Hot Topics in Operating Systems (HotOS)}}
\confname{\oopsla}{OOPSLA}{{Proceedings of the ACM SIGPLAN Conference on Object-Oriented Programming, Systems, Languages, and Applications (OOPSLA)}}
\confname{\ndss}{NDSS}{{Proceedings of the Network and Distributed System Security Symposium (NDSS)}}
\confname{\oakland}{Oakland}{{Proceedings of the IEEE Symposium on Security and Privacy (Oakland)}}
\confname{\ispass}{ISPASS}{Proceedings of the IEEE International Symposium on Performance Analysis of Systems and Software (ISPASS)}

\confname{\europar}{{E}uro{P}ar}{{Proceedings of the European Conference on Parallel Programming ({E}uro{P}ar)}}

\confname{\sigcse}{{SIGCSE}}{{Proceedings of the ACM SIGCSE Technical Symposium on Computer Science Education (SIGCSE)}}

\confname{\ccs}{{CCS}}{{Proceedings of the ACM Conference on Computer and Communications Security (CCS)}}

\confname{\veeconf}{{VEE}}{{Proceedings of the International Conference on Virtual Execution Environments (VEE)}}

\confname{\lisa}{{LISA}}{{Proceedings of the Large Installation System Administration Conference (LISA)}}
\confname{\scool}{SCOOL}{{Proceedings of the Workshop on Synchronization and Concurrency in Object-Oriented Languages (SCOOL)}}
\confname{\cgo}{CGO}{{Proceedings of the International Symposium on Code Generation and Optimization (CGO)}}
\confname{\dsn}{{DSN}}{Proceedings of the International Conference on Dependable Systems and Networks (DSN)}
\confname{\sac}{{SAC}}{{Proceedings of the ACM Symposium on Applied Computing (SAC)}}

\confname{\cluster}{{IEEE Cluster}}{{IEEE International Conference on Cluster Computing}}
\confname{\hotcloud}{HotCloud}{{Proceedings of the USENIX Workshop on Hot Topics in Cloud Computing (HotCloud)}}
\confname{\wlpe}{WLPE}{{Workshop on Logic-based methods in Programming Environments (WLPE)}}
\confname{\icce}{IC2E}{{IEEE Conference on Cloud Engineering (IC2E)}}
\confname{\apsys}{APSys}{{ACM Asia-Pacific Workshop on Systems (APSys)}}
\confname{\systor}{SYSTOR}{{ACM International Systems and Storage Conference (SYSTOR)}}
\confname{\hotstorage}{HotStorage}{{USENIX Workshop on Hot Topics in Storage and File Systems (HotStorage)}}
\confname{\issre}{ISSRE}{{IEEE International Symposium on Software Reliability Engineering (ISSRE)}}
\confname{\assets}{ASSETS}{{International ACM SIGACCESS Conference on Computers and Accessibility (ASSETS)}}
\confname{\chiconf}{CHI}{{ACM CHI Conference on Human Factors in Computing Systems (CHI)}}
\confname{\dsc}{DSC}{{IEEE Conference on Dependable and Secure Computing (DSC)}}
\confname{\mobicom}{MobiCom}{{International Conference on Mobile Computing and Networking (MobiCom)}}
\confname{\mobisys}{MobiSys}{{ACM International Conference on Mobile Systems, Applications, and Services (MobiSys)}}
\confname{\rtns}{RTNS}{{International Conference on Real-Time Networks and Systems (RTNS)}}
\confname{\systex}{SysTEX}{{Workshop on System Software for Trusted Execution (SysTEX)}}
\confname{\sfma}{SFMA}{{Workshop on Systems for Multi-core and Heterogeneous Architectures (SFMA)}}

\newcommand{\gcbench}{GCBench}
\newcommand{\rbench}{RBench}
\newcommand{\redis}{Redis}

\newcommand{\tabincell}[2]{\begin{tabular}{@{}#1@{}}#2\end{tabular}}

\newcommand{\us}{$\mu$s\xspace}

\ifx \oscardebug \undefined
\newcommand{\fixmeli}[1]{{}}
\newcommand{\fixmets}[1]{{}}
\newcommand{\fixmedp}[1]{{}}
\newcommand{\fixmetz}[1]{{}}
\newcommand{\fixmevd}[1]{{}}
\newcommand{\fixmebj}[1]{{}}
\newcommand{\fixmedm}[1]{{}}
\newcommand{\fixmees}[1]{{}}
\newcommand{\fixmemv}[1]{{}}
\newcommand{\fixmeef}[1]{{}}
\else
\newcommand{\fixmeli}[1]{{\bf\textcolor{red}{ [ FIXME Li: #1 ]}}}
\newcommand{\fixmets}[1]{{\bf\textcolor{red}{ [ FIXME Ts: #1 ]}}}
\newcommand{\fixmedp}[1]{{\bf\textcolor{red}{ [ FIXME dP: #1 ]}}}
\newcommand{\fixmetz}[1]{{\bf\textcolor{red}{ [ FIXME TZ: #1 ]}}}
\newcommand{\fixmevd}[1]{{\bf\textcolor{violet}{ [ FIXME VD: #1 ]}}}
\newcommand{\fixmebj}[1]{{\bf\textcolor{blue}{ [ FIXME BJ: #1 ]}}}
\newcommand{\fixmedm}[1]{{\bf\textcolor{green}{ [ FIXME DM: #1 ]}}}
\newcommand{\fixmees}[1]{{\bf\textcolor{violet}{ [ FIXME ES: #1 ]}}}
\newcommand{\fixmemv}[1]{{\bf\textcolor{red}{ [ FIXME MV: #1 ]}}}
\newcommand{\fixmeef}[1]{{\bf\textcolor{red}{ [ FIXME EF: #1 ]}}}
\fi

\newcommand{\intel}[0]{%
	\ifx \intelregistered \undefined%
		Intel\textsuperscript{\textregistered}%
		\def \intelregistered{}%
	\else
		Intel%
	\fi}

\newcommand{\amd}[0]{%
	\ifx \amdregistered \undefined%
		AMD\textsuperscript{\textregistered}%
		\def \amdregistered{}%
	\else
		AMD%
	\fi}

\newcommand{\callout}[1]{\vspace{10pt}\noindent\hspace{5pt}\fcolorbox{black}{cyan!20}{\parbox{3.1in}{{#1}}}\vspace{0pt}}

\begin{document}

\title{Adaptive and Efficient Dynamic Memory Management for Hardware Enclaves}


\author{\Large
	Vijay Dhanraj$^{\ast}$,
	Harpreet Singh Chawla$^{\dagger}$,
	Tao Zhang$^{\ddagger}$,
	Daniel Manila$^{\ddagger}$,\\
	Eric Thomas Schneider$^{\ddagger}$,
	Erica Fu$^{\ddagger}$,
	Mona Vij$^{\ast}$,
	Chia-Che Tsai$^{\dagger}$,
	Donald E. Porter$^{\ddagger}$}
\affiliation{$^{\ast}$
  \institution{Intel Corporation}
  \country{}
}

\affiliation{
  \institution{{$^{\ddagger}$The University of North Carolina at Chapel Hill}}
  \country{}
}

\affiliation{
  \institution{$^{\dagger}$Texas A\&M University}
  \country{}
}

\maketitle
\pagestyle{plain}

\subsection*{Abstract}

The second version of \intel{} Software Guard Extensions (\intel{} SGX), or SGX2,
adds dynamic management of enclave memory and threads.
The first version required the address space and thread counts
to be fixed before execution.
The Enclave Dynamic Memory Management (EDMM) feature of SGX2
has the potential to lower launch times and overall execution time.
Despite reducing the enclave
loading time by 28--93\%,
straightforward EDMM adoption strategies
actually slow execution time down by as much as 58\%.

Using the Gramine library OS as a representative enclave runtime environment,
this paper shows how to recover EDMM performance.
The paper explains how implementing mutual distrust between the OS and enclave increases
the cost of modifying page mappings.
The paper then describes and evaluates a series of optimizations on application benchmarks, showing
that these optimizations effectively eliminate the overheads of EDMM while retaining EDMM's performance
and flexibility gains.

\section{Introduction}

\intel{} SGX~\cite{intelsgx} is a powerful building block for protecting application code and data on a remote
system, such as cloud.
Specifically, an application can create a private region 
called an {\em enclave},
wherein the hardware protects sensitive code and data from system software, including
the OS and hypervisor(s).
The hardware encrypts enclave memory using a key known only to the CPU,
thereby protecting the enclave from unauthorized access
by system software and, in some configurations, via physical access (\S\ref{sec:background}).
SGX is particularly beneficial for securing cloud applications,
and multiple cloud providers offer SGX-protected platforms~\cite{SGXproduct,microsoft-sgx-vms,ibm-datashield,ibm-sgx,alibaba-sgx,alibaba-sgx2}.

SGX is an ambitious attempt at hardware supporting mutual distrust between system and application software.
However, \intel{} SGX version 1 ({\bf SGX1})
imposes constraints on runtime management of enclave virtual memory, leading to performance overheads as high as 58\% (\S\ref{sec:baseline:eval}).
Specifically, the enclave cannot dynamically add, remove, or protect virtual pages during execution.
The SGX driver can swap physical pages for an enclave, but the enclave's virtual memory layout 
must be set {\em before} execution.
The startup time for an enclave is directly correlated with enclave size, creating an unsavory trade-off:
a small enclave may run out of heap space, but has a fast
startup time;
in contrast, a large enclave has high startup times, but may over-allocate heap space.
These limitations negatively affect the utility of SGX1 in practical deployments.


\intel{} SGX version 2 ({\bf SGX2}) addresses these limitations with
Enclave Dynamic Memory Management (EDMM)~\cite{McKeen2016}.
SGX2 adds new instructions
that the SGX driver can use to add, remove, and change permissions of virtual pages in an enclave.
In order to maintain the integrity of the enclave, the enclave must explicitly accept any allocation, deallocation, or modification of virtual memory before the change will be accepted into the TLB.
In addition, newly allocated pages are zeroed by the hardware, as are deallocated pages.

{\bf Why does EDMM matter?}
Dynamic memory management is a fundamental necessity for any application that cannot statically predict its memory requirements,
such as sizing its heap.  Similarly, some applications need the ability to remove or change a mapping,
such as to read-protect static data after initialization.
Without EDMM, an application must overestimate its heap size
and over-permission the heap's pages as readable, writable, and executable.
Overestimating the heap size increases the enclave loading time in order to
map and
measure each page.
This issue is particularly salient for a compatibility layer, such as a library OS that implements system APIs
such as  {\tt brk}, {\tt mmap}, and {\tt mprotect},
as well as any userspace runtime environment that dynamically manages memory, such as a language runtime.



{\bf Why is designing an efficient EDMM scheme challenging?}
Intuitively, one would expect EDMM use to be a strict win: eliminate over-provisioned
memory resources, and reduce application startup time spent on measurement of heap pages that may never be used.
And indeed, using EDMM improves startup time.
However, straightforward use of EDMM makes runtime application performance {\em worse} in many cases.
For SGX2, we evaluate two separate strategies:
one strategy immediately maps pages into the enclave when the applications makes a call such as {\tt mmap};
the second strategy, proposed but not evaluated by Xing et al.~\cite{Xing2016},
is demand allocation, which allocates upon the first access to a virtual page.
The results of an experiment running \gcbench{}, a garbage collection
benchmark, shows that
both EDMM designs cause significant slow down to the execution time, up to 41\%
and 58\%, compared to just using SGX1.
More details of the result are in \S\ref{sec:baseline:eval}.



The key intuition behind this result is that modifying memory mappings 
is more expensive
in an enclave than in a normal process, due to the additional context switching required
for the enclave to accept the changes.
Adding an enclave mapping requires at least three context switches, and at least five if the mapping is added using demand paging (\S\ref{sec:background}).
Removing an enclave mapping is even more expensive, requiring at least
{\em nine} context switches to ensure that the TLB entry is properly flushed.
On our test machine, the cost of a demand fault on a newly mmap-ed area in a normal process is about 8~\us, whereas creating an enclave page mapping takes around 30~\us on SGX2. 

{\bf Primary goals and findings:}
This paper investigates and addresses these overheads.
We evaluate real-world applications on Gramine (formerly Graphene)~\cite{Tsai2017, tsai14graphene},
as a representative compatibility layer for SGX. The {\bf contributions} of this paper are as follows:
\begin{compactitem}
\item An analysis on how straightforward use of EDMM {\em increases} application execution time, as does the optimization suggested in previous work~\cite{Xing2016}.
\item A series of optimizations for EDMM in Gramine and SGX that {\em remove} the overheads while retaining the space efficiency gains of EDMM.
\item A thorough evaluation of these optimizations using both microbenchmarks and applications. These optimizations significantly enhance application performance, making EDMM in Gramine comparable or better to static allocation, while greatly improving application startup time.
\end{compactitem}

\section{Background}
\label{sec:background}

This section introduces background on SGX and
the Enclave Dynamic Memory Management (EDMM) feature of SGX2,
and relates this model to the more recent VM-based trusted-execution model
of Intel TDX and AMD SEV.





Hardware trusted execution environments (TEEs),
such as \intel{} SGX~\cite{intelsgx} and AMD SEV~\cite{amd-sev}, protect
security-sensitive computation and data from system-level attackers, including malware, OS rootkits, and sometimes even physical attackers.
TEEs
isolate
the execution of a program, including CPU registers and memory,
from the host operating system and other software.
The CPU ensures integrity,
and can generate attestation reports as proofs of integrity to remote entities.
When a program runs inside a TEE, the program's memory
is protected by hardware, using encryption~\cite{intelsgx, amd-sev, arm-secureip} or access control~\cite{keystone-enclave}, preventing a system attacker from accessing program memory.
Currently, hardware TEEs have been widely applied for data-intensive computation~\cite{vc3, Kuccuk16, priebe2018enclavedb, Safebricks, ShieldStore, Vessels}, control-plane software~\cite{SecureKeeper, Pires16}, and privacy-preserving systems~\cite{Pires16, Kim17, Firch17, Oblix, ObliDB}.

This paper focuses on \intel{} SGX, one of the earliest hardware
TEEs in widespread production. 
A hardware {\em enclave} created by SGX is a protected virtual memory range
within a process's address space.
In SGX version 1,
the CPU does not allow changes to the enclave's virtual address space
after initialization of the enclave.
The only exception is swapping: the
untrusted host OS may update the page table to unmap or remap an enclave page,
but this involves a more complex check that the contents did not change
while the page was unmapped.
Swapped pages are encrypted by the hardware.
Early versions of SGX limited the physical memory for all enclaves on a machine to 128MB;
with the addition of \intel{} Total Memory Encryption (TME)~\cite{tme-whitepaper}, SGX now supports enclaves as large as 1 TB.

\begin{table}[t!]
\centering
\small
\begin{tabular}{lp{2.4in}} \toprule
{\bf Leaf Func.}& \multicolumn{1}{c}{\bf Description } \\ \midrule
  \multicolumn{2}{c}{{\bf System-tier ({\tt ENCLS})}} \\ \midrule

EAUG & Add a zeroed virtual page to an enclave. \\
EMODT & Modify the page type of a virtual page. \\
EMODPR & Reduce access permissions of a virtual page. \\
\midrule
\multicolumn{2}{c}{{\bf User-tier ({\tt ENCLU})}} \\ \midrule
EACCEPT & Accept addition or changes of a virtual page made by the OS\\
\multicolumn{1}{p{.9cm}}{EACCEPT\-COPY} & Copy an existing virtual page into an {\tt EAUG}'ed page and accept the {\tt EAUG}'ed page into the enclave. \\
EMODPE & Extend access permissions of a virtual page.\\
\bottomrule
\end{tabular}
\caption{Summary of EDMM leaf functions.}
\label{Tab: edmm_instr}
\end{table}

\subsection{Enclave Dynamic Memory Management (EDMM)}
\label{subsec:EDMM}

\intel{} SGX version 2 (SGX2) introduced
Enclave Dynamic Memory Management (EDMM).
EDMM includes
new {\em leaf functions} to SGX's {\tt ENCLS} (system-tier) and {\tt ENCLU} (user-tier) instructions,
to add, remove, and protect a virtual page within an enclave.
In SGX, {\tt ENCLS} is only called inside the kernel, while {\tt ENCLU} is called from userspace, either within or outside an enclave.
The functionality of {\tt ENCLS} and {\tt ENCLU} is determined by an extra opcode (given by the {\tt EAX} register), to indicate a leaf function.
Table~\ref{Tab: edmm_instr} lists the new leaf functions added for EDMM.
For simplicity, this paper describes these leaf functions as instructions.

System-tier leaf functions ({\tt ENCLS}) require agreement from the enclave software,
since the host OS is not trusted to always provide the correct parameters.
A benign OS is
responsible for adding, removing, or protecting a virtual page in an enclave,
as well as updating the page table accordingly.
For example, to make an initially unmapped virtual page available, the OS needs to call {\tt EAUG} with a physical page from the {\bf EPC} (Enclave Page Cache, or physical pages reserved for use in enclaves).
The OS subsequently maps the virtual page to the physical page in the page table, with the corresponding access permissions.
The enclave must then approve the changes using {\tt EACCEPT} to make the changes take effect.
Note that to extend the access permissions of a virtual page, the enclave does not need the OS to intervene and can directly request the change using {\tt EMODPE}, assuming that the page table is more permissive for this virtual address.

Anecdotally, the introduction of EDMM is useful for libraries or unikernels~\cite{Arnautov2016, tsai14graphene, Baumann2014} for porting legacy applications into enclaves:
(1) EDMM allows dynamic allocation or deallocation of virtual pages after enclave launch. Without EDMM, most enclaves will have to overpopulate virtual memory,
increasing enclave startup time.
(2) EDMM allows an enclave to dynamically and securely change page permissions after enclave launch, which is crucial for implementing systems APIs like {\tt mprotect}, as well as supporting the process of ELF loading.
Without EDMM, a library OS needs to {\em unsafely} make the entire heap readable, writable, and executable,
because {\tt mprotect} does not function after enclave launch.
(3) EDMM allows dynamically changing a virtual page to a Thread Control Structure (TCS) page, which is needed for thread creation.
Without EDMM, the number of enclave threads must be determined statically, and multithreaded enclaves may have to overestimate the number of threads since TCS pages cannot be added afterwards.

In this paper, we focus on optimization of the system flow for allocation and
deallocation of virtual pages and kernel threads. We leave optimization of
thread creation and changing page permissions to future work.

\subsection{VM-based TEE Memory Management}
\label{subsec:vm-based}



Another trend of trusted execution environment (TEE)
is to place an entire virtual machine (VM) in a hardware-protected domain, in which the sensitive application can be fully supported by a trusted guest OS.
For example, AMD SEV-SNP \cite{amd-sev,amd-sev-snp} isolates VM state from the hosting hypervisor, with the VM's associated physical memory encrypted by the CPU, and changes to guest-to-host memory mapping being detectable.
\intel{} Trusted Domain Extensions (TDX) \cite{tdx} adopts a similar strategy, by including the entire DRAM as the EPC (Total Memory Encryption) and protecting the extended page table (EPT) from the host.

For these VM-based TEEs,
the virtual memory of the sensitive application(s) in the VM
is fully managed by the guest OS,
which also manages a pool of physical pages with hardware encryption.
Unlike SGX, a VM-based TEE does not require intervention of the host to extend or shrink the virtual memory space of applications.
But because both VMs and EDMM-enabled enclaves must dynamically add
and remove physical pages from a TEE,
they require a similar flow to SGX's EDMM.
For example, in TDX, a protected VM
needs the host OS to issue the same {\tt EAUG} instruction to add a physical page, and then the VM needs to approve the change using {\tt EACCEPT}.
This suffers a similar as SGX2's EDMM,
although it may occur less frequently since some changes,
such as changing page permissions or remapping a page,
can be made by
the guest OS
without involving the host OS.

\section{Baseline EDMM Performance}


We study the overheads of EDMM in the Gramine library OS~\cite{Tsai2017}, as a representative example
of a ``lift-and-shift'' framework for running applications in SGX.
This section begins by describing baseline virtual memory management
support in Gramine, followed by our straightforward implementation
of EDMM support in Gramine.
We implemented all operations synchronously. \fixmedp{exitless would be a short-list goal}

\subsection{Threat Model}
This work follows a common threat model for SGX applications.
The only trusted components are the CPU(s), remote attestation services, and the code and data
running inside the enclave.
For remote attestation, we trust a special enclave and its containing software, called {\tt aesmd}, provided by Intel.
The hardware outside of the CPU package, the hypervisor, OS,
and other code outside of the enclave are untrusted.
The recent {\em Scalable SGX} foregoes integrity protection against physical attacks, such as a memory interposer, but does
ensure confidentiality against physical tampering.
Our prototype, based on Gramine, uses the in-kernel SGX driver, but does not trust this driver.
Denial-of-service, cache-based side-channels, and controlled channel attacks~\cite{Xu2015}
are out of the scope of this work.

\subsection{Experimental Setup}
We use a desktop, with a 144-core 2.40 GHz Intel(R) Xeon(R)
Platinum 8360Y CPU with a 108 MB L3 cache and hyperthreading enabled, 248 GB RAM,
an approximately 4 GB EPC, and
a Samsung 970 EVO 500 GB SSD. The host OS is Ubuntu 20.04.6 LTS, with Linux
kernel 6.7.0 and the in-kernel SGX driver.
Our kernel includes a patch to the SGX driver to support an optimization described later (\S\ref{sec:approach:range});
this patch does not affect the behavior of the SGX driver when the optimization is not used, as in these tests.

Unless otherwise noted, results are the average of 10 runs,
and error bars represent 95\% confidence intervals.

\vspace{0.5em}
\noindent
{\bf Application Workloads.}
We use three applications to evaluate EDMM performance: \gcbench{}, \rbench{}, and \redis{}.
\gcbench{}~\cite{gcbench} exercises the Python (version 3.8.10) garbage collector
by allocating and reclaiming representative data-structures,
such as a set of balanced binary trees and a large array of floating point values.
We use version 2.5 of \rbench{}~\cite{rbench}, which tests a variety of scientific computing tasks, such as matrix computation and calculating Fibonacci numbers, using version 3.6.3 of R.
Lastly, \redis{}~\cite{redis} uses Yahoo! Cloud Serving Benchmark (YCSB) \cite{ycsb} to test the performance of \redis{} (version 6.0.5),
an in-memory key-value database.
For \rbench{} and \gcbench{}, we collect the end-to-end execution time using
the {\tt time} command.
YCSB on \redis{} reports throughput across six sub-workloads.
We size enclaves for running \rbench{} benchmark at 2 GB, and \gcbench{} benchmark, \redis{} server at 512 MB, based on their memory usage.





\subsection{Baseline 1: Static Allocation ({\sf static})}

Gramine was originally designed for SGX1,
on which the enclave's virtual memory layout is determined statically.
Because of this limitation, Gramine on SGX1 cannot implement
memory-related system calls, including {\tt brk}, {\tt mmap}, {\tt munmap}, and {\tt mprotect}.
Although the virtual memory space is statically configured, Gramine does still dynamically
assign allocated pages to purposes such as the heap,
using Gramine-internal bookkeeping.






\begin{figure}[t!]
  \includegraphics[width=.25\textwidth]{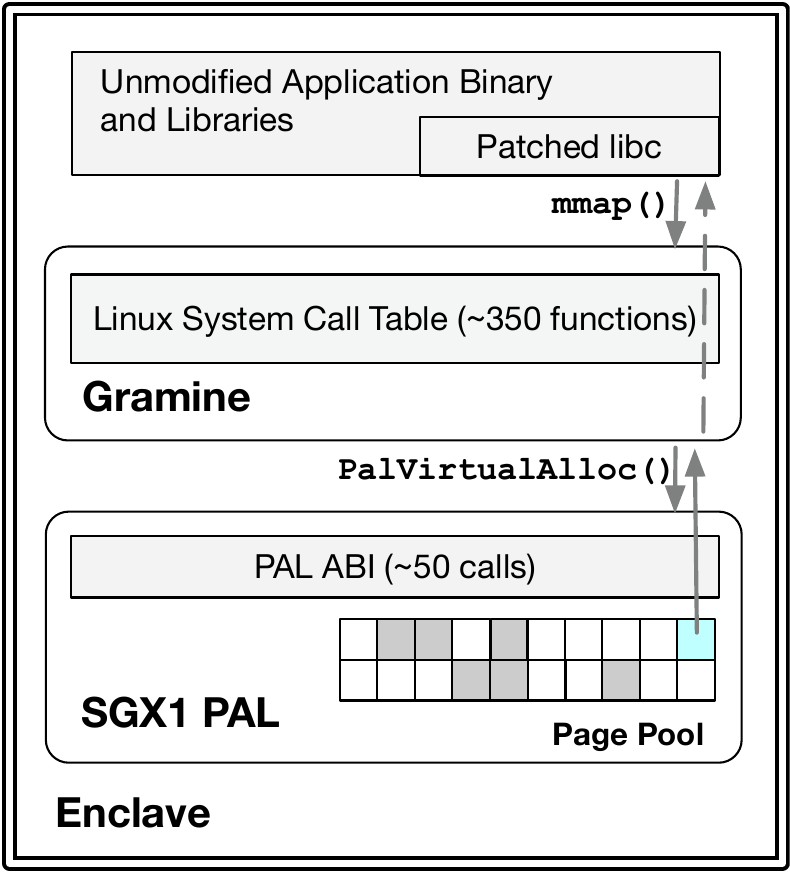}
  \vspace{-10px}
  \caption{The Gramine architecture on SGX1. 
  	The PAL (platform adaptation layer) internally manages a page pool to return statically allocated, free virtual pages (in cyan) to {\tt mmap}.}
  	 \label{fig:gramine-arch-mmap}
\end{figure}

Gramine implements enclave management memory logic in its {\bf Platform Adaptation Layer (PAL)}.
Gramine implements one PAL per supported host platform (e.g., Linux on SGX);
the PAL abstracts and encapsulates host-specific differences, so that the library OS can run on top of any PAL without modification.
The three PAL APIs for memory management include:
{\tt Pal\-Virtual\-Memory\-Alloc}, {\tt Pal\-Virtual\-Memory\-Free}, and {\tt PalVirtual\-Memory\-Pro\-tect}.
Figure~\ref{fig:gramine-arch-mmap} shows the architecture of Gramine and its memory management mechanisms.
%

On SGX1, Gramine implements these APIs
using a {\bf page pool} to manage the heap.
Gramine pre-allocates a range of virtual pages within the enclave as the page pool.
The page pool is used to serve subsequent requests for pages, including mapping files or
extending the heap.
Because these pages cannot change permission dynamically, they are mapped readable, writable, and executable.
Because most dynamic linking happens after enclave launch in Gramine,
only the PAL loader binary is mapped read-only; all other binaries are in
 writable inside the enclave.
Gramine implements {\tt PalVirtMemoryProtect} as a no-op on SGX, since SGX1 does not allow changing page permissions during runtime.
In the experimental results, we refer to this design as {\sf static}.

\subsection{Baseline 2: Basic EDMM Support ({\sf edmm})}
\label{sec:baseline:naive}

\begin{figure}[t!]
	\includegraphics[width=.4\textwidth]{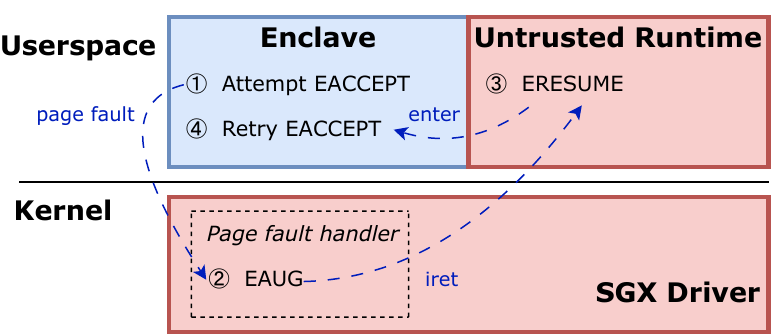}

	\caption{The current system flow for dynamically adding a page to an SGX enclave with EDMM.  Trusted software is shaded blue. 
	  Compared to allocating a page to a normal process, this requires an additional context switch, back into the enclave, and that the enclave accepts the new mapping. }\label{fig:eaug-basic}
\end{figure}

We first describe a straightforward approach to adopting EDMM in Gramine.
We use the same page pool abstraction
as described previously,
but rather
than map all of these pages at enclave launch, we instead map them in response to
the first use by an enclave-level, emulated system call in Gramine.
For instance, when an enclave application issues an {\tt mmap} call to Gramine,
Gramine will select an unmapped page from the page pool to return to the application.
The application could trigger a demand fault by simply reading or writing to the page,
which would be serviced by the SGX kernel driver.
The SGX kernel driver then creates the mapping (using {\tt EAUG}), and returns to the enclave, which in turn
accepts the new mapping, exits the enclave, and then re-enters the enclave to resume execution.
This additional enclave exit is an SGX hardware requirement.


We adopt an optimization  proposed, but not evaluated, by Xing et al. \cite{Xing2016} (their \S5.2),
that avoids the additional exit between accepting the mapping and resuming execution.
The trick is to have the {\tt EACCEPT} be the instruction that triggers the fault, so that
the page is accepted while resuming execution.
This is illustrated in Figure~\ref{fig:eaug-basic}.
When an enclave issues an {\tt EACCEPT} instruction on an unmapped virtual address, this exits the enclave and raises a page fault in the kernel.
Then, the SGX driver will map the faulting address using {\tt EAUG}.
When the process resumes execution, it will pass through the untrusted runtime to transition back
into the enclave;
then, the enclave will sucessfully retry the {\tt EACCEPT}.
Gramine implements this by synchronously calling {\tt EACCEPT} on each virtual page mapped by {\tt PalVirtMemoryAlloc} (called by either {\tt mmap} or {\tt brk}).
In the experiment results, we refer to this design as {\sf edmm}.

\fixmedp{One assumption from the HASP paper to verify/refute: an ocall is more expensive than a page fault?  Not obvious to me why that is true}




\begin{figure}[t!]
  \includegraphics[width=.5\textwidth]{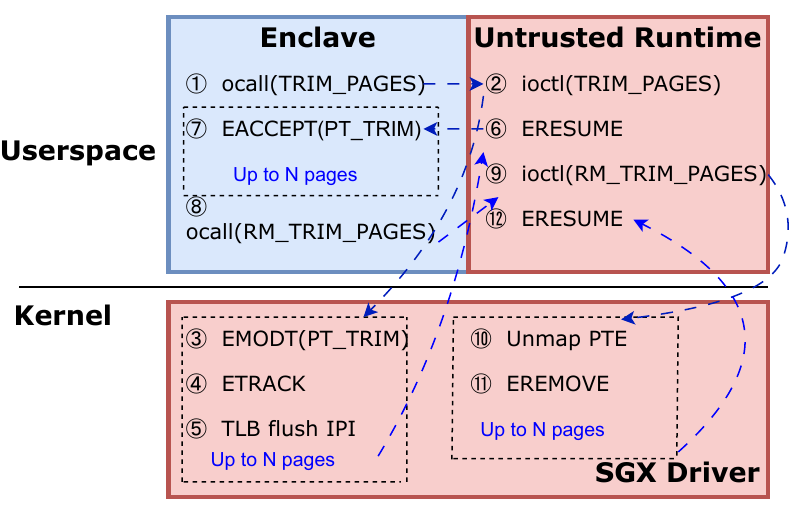}
  		\vspace{-20pt}
  \caption{The system flow for removing a virtual page mapping from an enclave.}\label{fig:naive-free}
\end{figure}

Removing a mapping in EDMM
uses a similar, but more complex system flow, in order to avoid race conditions
with stale TLB mappings.
Figure~\ref{fig:naive-free} illustrates the system flow recommended by the Intel manual,
which is implemented by the SGX driver.
To remove a virtual page, the enclave first issues an {\tt ocall} to exit the enclave, and then uses {\tt ioctl} to get the SGX driver to transition each page in the range into the {\tt TRIM} (or pending removal) state.
The driver also blocks creation of any new mapping at the page (using {\tt ETRACK}) and shoots down any cached TLB mappings on each CPU core.
Then, the flow returns to the userspace and re-enters the enclave to call {\tt EACCEPT}.
Next, the enclave issues an {\tt ocall} again to exit the enclave and then uses {\tt ioctl} to officially call {\tt EREMOVE} on each virtual page.

\fixmevd{Cost of an AEX/per EPC page will be good to have}


\subsection{Baseline 3: Demand Allocation ({\sf edmm+demand})}
\label{subsec:demand_paging}
\label{sec:baseline:demand}

\begin{figure}[t!]
	\includegraphics[width=.4\textwidth]{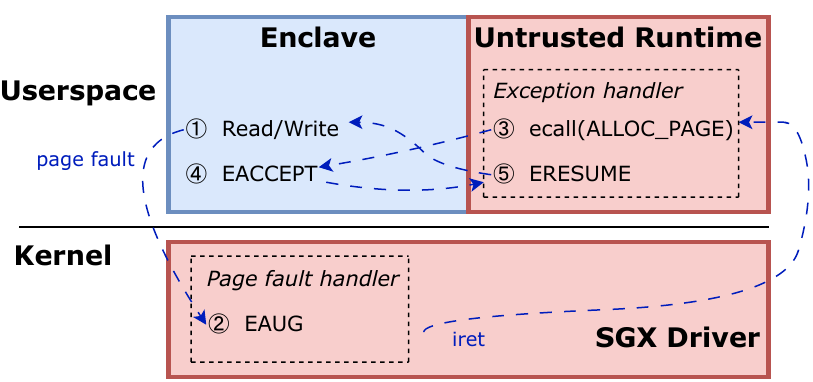}
	\caption{System flow of creating a new enclave virtual memory mapping via demand allocation, involving five context switches among the untrusted kernel, Gramine's untrusted runtime, and the enclave.  Trusted components are in blue, untrusted in pink.}\label{fig:demand-proc}
\end{figure}

\begin{figure*}[t!]
  \begin{minipage}{\textwidth}
		\includegraphics[width=.40\textwidth, height=6cm, keepaspectratio]{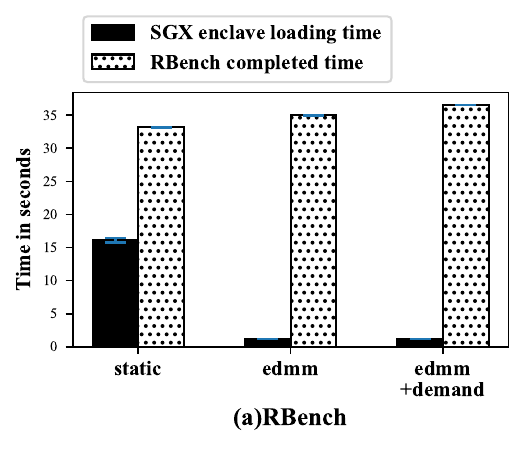}
		\hspace{5px}
		\includegraphics[width=.48\textwidth, height=6cm, keepaspectratio]{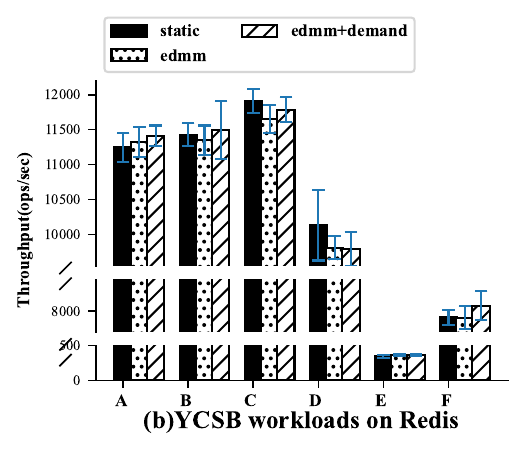}
	\end{minipage}\\
	\vspace{5px}
	\begin{minipage}{\textwidth}
		\includegraphics[width=.40\textwidth, height=6cm, keepaspectratio]{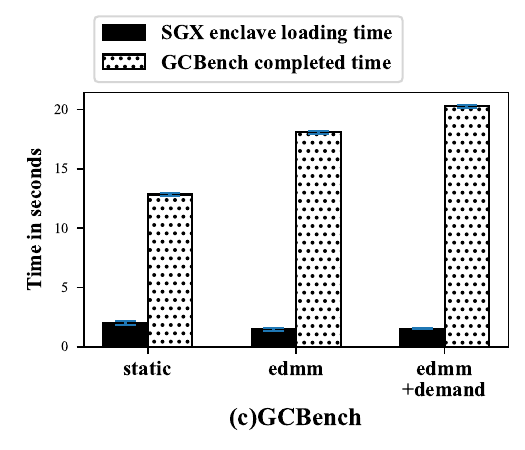}
		\hspace{5px}
		\includegraphics[width=.48\textwidth, height=6cm]{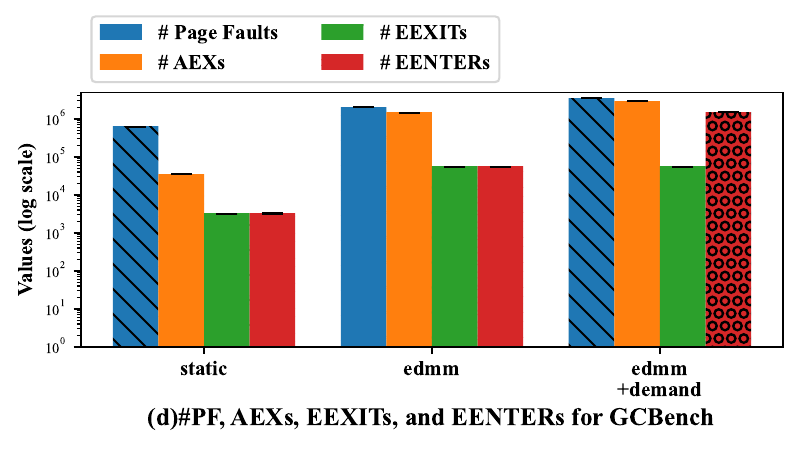}
	\end{minipage}
	\caption{Benchmarking of three application workloads: (a) RBench; (b) Redis; (c) GCBench, and (d) the numbers of page faults, AEXs, EEXITs, and EENTERs during the GCBench execution.
		Each set of results is collected on static allocation ({\sf sgx1}), basic EDMM support ({\sf edmm}), and EDMM with demand allocation ({\sf edmm+demand}). Lower is better.
	}
	\label{fig:baselines}
\end{figure*}

Demand allocation is a standard optimization in memory management.
In prior work, Xing et al.~\cite{Xing2016} propose, but do not implement
or evaluate, a design for demand allocation in SGX.
They propose an {\em implicit region}
as an abstraction within the enclave's address space, wherein a page fault
is assumed to be an implicit allocation request to the kernel.
The enclave would communicate these regions to the kernel and
SGX driver in advance of the page faults.

We adapt this design to Gramine by treating the entire enclave page
pool as the implicit region.
When the enclave launches, the page pool is initially unmapped.
Just as with the baseline EDMM design, as the application or library OS requests mappings, Gramine logically allocates regions of the page pool.
The key difference is that each page in the
page pool now has a state, mapped or unmapped, which Gramine tracks
with an additional bit per page.
The first time a page from the page pool is allocated to an application,
it will be in the unmapped state.  Once the page is faulted in,
Gramine transitions it to the mapped state, where the page remains until
the mapping is removed.


A key distinction compared to Baseline 2 is that the faulting
instruction is no longer {\tt EACCEPT}; this means that
the enclave must provide an exception handler to issue the {\tt EACCEPT}.
Figure~\ref{fig:demand-proc} illustrates the revised system flow.
When the enclave accesses a virtual address that is not yet mapped,
a page fault is raised and the kernel handles the fault.
The SGX driver maps the page using {\tt EAUG}
and returns to untrusted, user-level runtime, which then re-enters the enclave in order to invoke the in-enclave exception handler.
The exception handler consults Gramine's internal bookkeeping
to confirm that the faulting address is in a virtual page that
the application requested and not yet been accepted.
Then the exception handler calls {\tt EACCEPT} on the virtual page
and marks the page as mapped.

After accepting the new mapping,
the enclave must again context switch back to the untrusted runtime and re-enter the enclave to resume execution.
The underlying issue is the hardware
needs to decrement a counter that tracks the number of register-saving regions used so far,
and this can only happen when resuming an enclave execution using {\tt ERESUME}.
Newer versions of SGX (including upcoming firmware updates for some existing chips) has a feature called AEX-Notify~\cite{aex-notify-wp}
that will allow an enclave exception handler to resume normal execution
without the additional context switches.
We will adopt and evaluate the impact of AEX-Notify in future work.
\fixmedp{We should do this.}

\subsection{Baseline Performance Evaluation}
\label{sec:baseline:eval}


We start with measuring the performance of three application workloads in Gramine on the three baselines:
static allocation, basic EDMM support, and EDMM with demand allocation.
Figure~\ref{fig:baselines} shows the enclave loading time and execution time of \rbench{} and \gcbench{},
as well as the throughputs of YCSB benchmarks on \redis{},
and the numbers of page faults, asynchronous exits (AEXs), enclave exits (EEXITs), and enclave enters (EENTERs) for \gcbench{}.
We omit these numbers for \rbench{} and \redis{} because they show a similar trend as the numbers for \gcbench{}.

The trends in Figure~\ref{fig:baselines} for loading time and execution time are consistent:
while basic EDMM support decreases the enclave loading time by 28--93\%, it increases the execution time by 5\% for \rbench{} and 41\% for \gcbench{} versus static configuration.
This performance degradation correlates with an increase in costly page faults
and other enclave crossings.

Not only does adding basic EDMM support harm performance, demand allocation
worsens the situation further.
Overall, demand allocation increases the execution time by 10\% for
\rbench{} and 58\% for \gcbench{};
worse, demand allocation does not reduce enclave loading time further
compared to basic EDMM support.
For \redis{}, demand allocation results in both gains and losses across different workloads.
These variations are relatively small, due to the minimal demand allocation occurring during individual requests.
Although demand allocation is an optimization in other contexts,
in the case of EDMM,
demand allocation requires
an additional, costly enclave entrance and exit to accept the new mapping,
which is not offset by any gains from delaying creation of the mapping.



\callout{\textbf{Insight}: EDMM's enclave loading time gains can be quickly offset by expensive enclave crossings required for dynamic memory management,
  causing a net slowdown in application performance.}

\section{Removing EDMM Overheads}


This section presents four optimizations
that reduce the overheads demonstrated in the previous section.
First, because it is cheaper to create static mappings at loading time than dynamic mappings,
we show how a modest pre-allocation strategy can reduce overheads without
over-provisioning space.
\fixmedp{Evaluation should really report some space data to show this trade-off}
Second, one can amortize the cost of dynamic mappings by
batching one set of enclave entrace and exits across multiple
{\tt mmap} requests.
Third, we observe that demand allocation often occurs on contiguous virtual pages, so we can proactively map neighboring pages in response to a demand fault.  
Finally, because unmapping is more expensive than mapping pages in an enclave, an asynchronous, lazy unmapping strategy can further reduce costs and
create opportunities for mapping reuse.
In order to understand the impact of each optimization, we evaluate the improvement over baseline Gramine (and the prior optimizations)
in each subsection.









\subsection{Optimization 1: Pre-allocation ({\sf +pre})}

\begin{figure*}[t!]
	\begin{minipage}{\textwidth}
		\includegraphics[width=.48\textwidth]{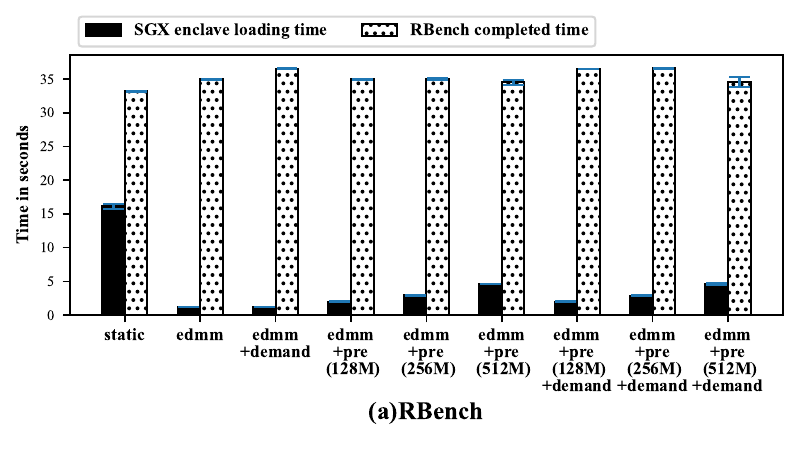}
		\hspace{5px}
		\includegraphics[width=.48\textwidth]{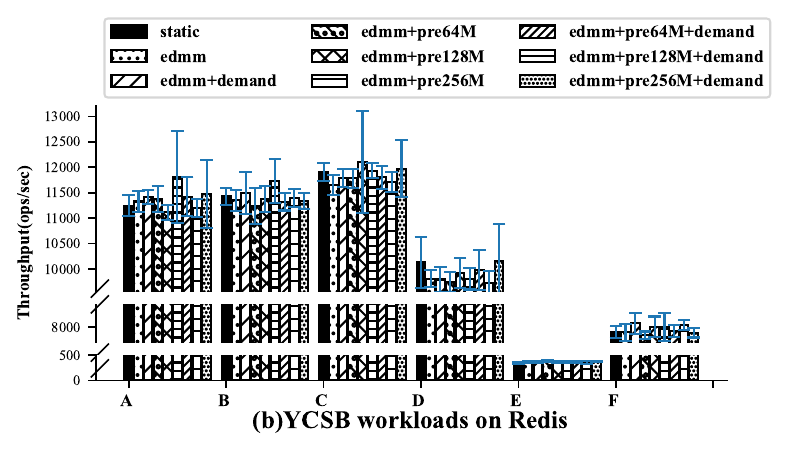}
	\end{minipage}\\
	\vspace{10px}
	\begin{minipage}{\textwidth}
		\includegraphics[width=.48\textwidth]{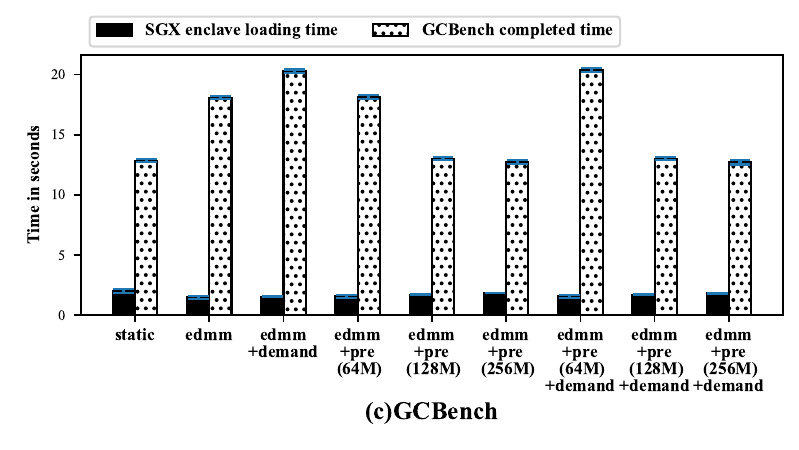}
		\hspace{5px}
		\includegraphics[width=.48\textwidth]{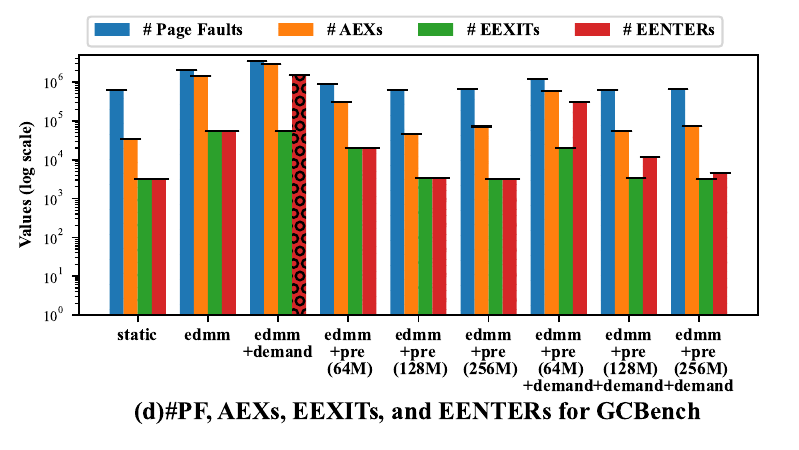}
	\end{minipage}
	\caption{({\bf Pre-allocation}) Benchmarking of three application workloads: (a) RBench; (b) Redis;  (c) GCBench, and (d) the numbers of page faults, AEXs, EEXITs, and EENTERs during the GCBench execution.
		Each set of results is collected on the three baselines ({\sf sgx1}, {\sf edmm}, and {\sf edmm+demand}), EDMM with different pre-allocation sizes: 64M ({\sf +pre(64M)}), 128M ({\sf +pre(128M)}),
    256M ({\sf +pre(256M)}), and 512M ({\sf +pre(512M)}), either with or without demand allocation ({\sf +demand}).
	}
	\label{fig:preallocation}
\end{figure*}

Given that it is faster to allocate memory at enclave launch than
during runtime, the first optimization is simply to pre-allocate
a reasonable starting size for the heap.
Although the number of pages used by an application can be input-dependent, one can often reliably predict
a minimum amount of expected dynamic memory usage, say based on the smallest input.

Our first optimization, named {\bf pre-allocation} (denoted as {\sf +pre}), allows the user to specify an initial page pool size
to fully allocate during enclave launch.
Gramine allocates an initial set of pages at launch, and
after the application exhausts this allocation, Gramine switches to dynamic allocation.
During the initial stage of the enclave execution,
Gramine has a fixed memory footprint,
mostly storing internal data structures of Gramine and loading application binaries.
Pre-allocation prevents frequent enclave exits or page faults until the application is fully loaded, as well as servicing some memory allocation requests from application itself.

We note that pre-allocating too much memory may have downsides:
first, the enclave potentially uses more
memory than needed, and, with a sufficiently large initial page pool size,
degenerates to the static configuration.
Second, as our evaluation shows,
pre-allocating memory increases enclave loading time.
If the space is going to be used, pre-allocation is more efficient than dynamic
allocation and this is a net win; if the space is unused, this is a needless
performance cost.
Worse, if the system is under memory pressure, needless
pre-allocation
can impact the performance of other running enclaves or even regular applications. We leave experiments on memory pressure for future work.
\fixmedp{Good idea to beef this paper up, tho}

\fixmedp{Is it possible to break the y-axis on the redis graphs so we can see a bit more detail?}

{\bf Evaluation.}
Figure~\ref{fig:preallocation} shows the impact of pre-allocation on our three application workloads: \rbench{}, \redis{}, and \gcbench{}.
We experiment with different preallocation sizes: 64M, 128M, 256M, and 512M.
We also test the optimization on two EDMM baselines: the basic EDMM support, and demand allocation.
For \rbench{}, pre-allocation with 512M brings down the cost of using EDMM to be on par with static allocation,
yielding only a 4\% slowdown in execution time but 3.5$\times$ faster enclave load time compared to static allocation.
For \gcbench{}, pre-allocation with 128M shows performance on par with static allocation regarding execution time, while also providing better enclave load time.
Contrast this with baseline EDMM overheads of 58\% with and 41\% without demand paging.
For \redis{}, pre-allocation with 256M improves almost all of the workloads over static allocation.
Workload A has improvement up to 4\%.

Fig.~\ref{fig:preallocation}(d) presents enclave entrances and exits
for \gcbench{}, showing a reduction in enclave crossings
that corresponds to the performance gains for pre-allocation.






\callout{\textbf{Insight}: Pre-allocating pages trades an increase in loading time for a larger reduction in execution time, provided the pages are used.  Pre-allocating unused pages lowers execution time compared to not mapping them.}

\subsection{Optimization 2: Batch Allocation ({\sf +batch})}
\label{sec:approach:range}






\begin{figure*}[t!]
  \begin{minipage}{\textwidth}
		\includegraphics[width=.40\textwidth, height=6cm, keepaspectratio]{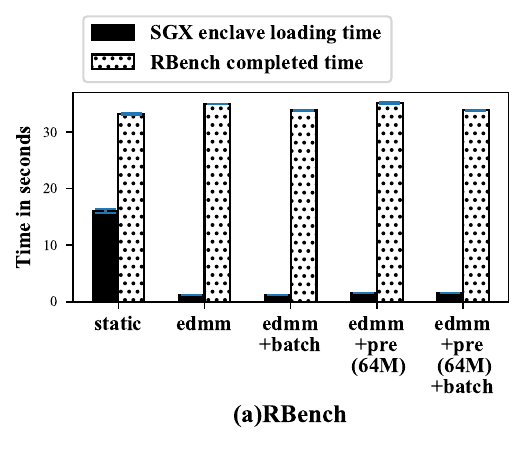}
		\hspace{5px}
		\includegraphics[width=.48\textwidth, height=6cm, keepaspectratio]{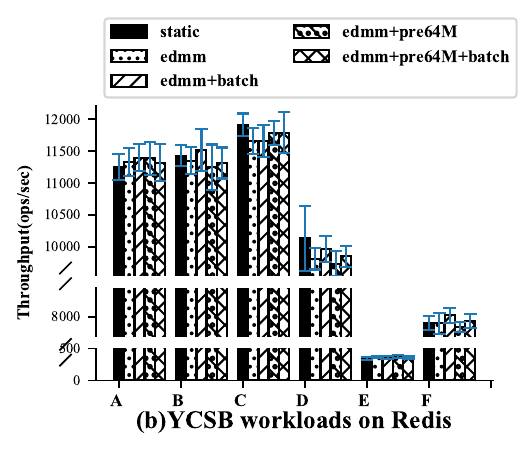}
	\end{minipage}\\
	\vspace{5px}
	\begin{minipage}{\textwidth}
		\includegraphics[width=.40\textwidth, height=6cm, keepaspectratio]{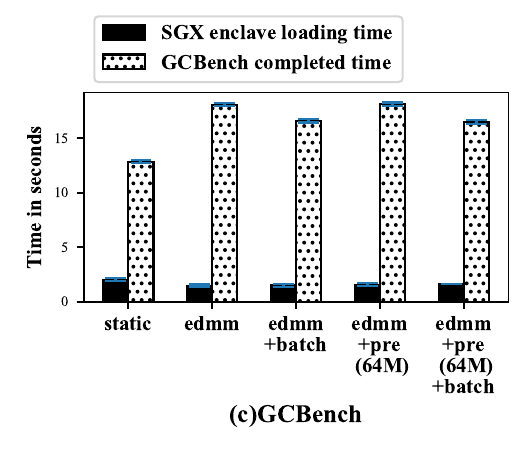}
		\hspace{5px}
		\includegraphics[width=.48\textwidth, height=6cm]{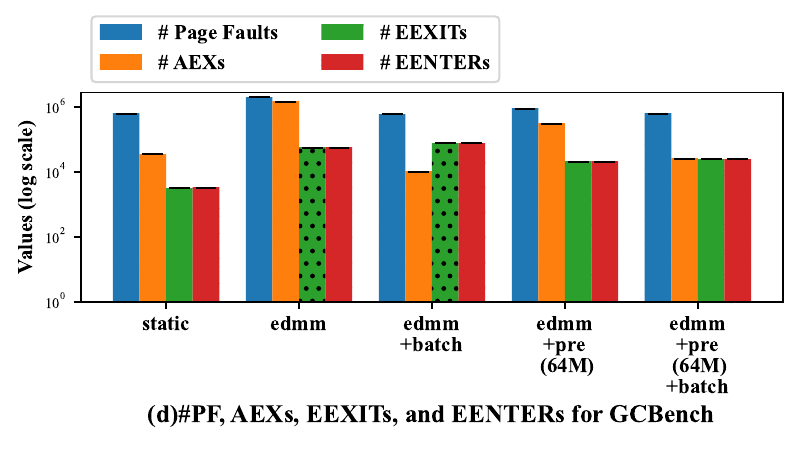}
	\end{minipage}
	\caption{({\bf Batch Allocation}) Benchmarking of three application workloads: (a) RBench; (b) Redis;  (c) GCBench, and (d) the numbers of page faults, AEXs, EEXITs, and EENTERs during the GCBench execution.
		Each set of results is collected on the three baselines ({\sf sgx1}, {\sf edmm}, and {\sf edmm+demand}), EDMM with only batch allocation ({\sf +batch}), EDMM with 64M pre-allocation ({\sf +pre(64M)}), either with or without batch allocation ({\sf +batch}).
	}
	\label{fig:batch}
\end{figure*}

A major source of runtime overhead in EDMM is the additional, synchronous context switch
for the enclave to approve each change to a page mapping.
Xing et al.~\cite{Xing2016} propose, but do not implement or evaluate, a solution that
amortizes one single round trip into the kernel over a virtually contiguous set of pages. We call this technique {\bf batch allocation} (denoted as {\sf +batch}).  
We note that the technique of batch allocation defined in this paper is specifically an optimization to the basic EDMM support.
A similar idea of ``batching'' can be applied to demand allocation, which we will discuss in \S\ref{subsec:cont_demand}.

Recall that, in basic EDMM support without demand paging,
when an application issues a multiple-page {\tt mmap},
Gramine allocates a region from the page pool.
Gramine then
issues an {\tt EACCEPT} instruction on each virtual page
in the newly mapped range,
which triggers a demand fault on each page.
This causes the kernel to issue the {\tt EAUG} instruction,
which then allows the {\tt EACCEPT} to succeed.
Put differently, this strategy incurs one enclave crossing and one system call
per page mapped.

Rather than demand fault these pages one at a time,
the key intuition of batch allocation is to amortize
one enclave and kernel crossing over an {\tt mmap}-ed range.
Before Gramine's {\tt mmap} implementation issues an {\tt EACCEPT}
instruction on each page, it first issues an {\tt ocall} to the untrusted runtime,
which in turn issues a special
{\tt madvise} system call to the SGX driver
on behalf of the enclave.
This {\tt madvise} call tells the kernel the location and size
of the mapping change, causing the driver to issue
a series of {\tt EAUG} instructions over the requested virtual address range.
Upon return to the enclave, the enclave can then issue a series of {\tt EACCEPT} instructions over the same range without further page faults.


This {\tt madvise} feature is not implemented in the Linux SGX driver yet.
We are working with the \intel{} team and plan to upstream a patch to Linux in the future.

In total, the batch optimization lowers the costs of dynamically modifying
page mappings from one enclave-kernel round trip per page to one per contiguous memory region.
From the application's perspective, there is no demand allocation in either the baseline or with this optimization; after an
{\tt mmap}, the returned region of virtual address space is fully mapped and usable.

\vspace{0.5em}
\noindent
{\bf Evaluation.}
Fig.~\ref{fig:batch} shows the impact of batch allocation on our three application workloads over the basic EDMM support with the pre-allocation optimization.
For simplicity, we pick the smallest pre-allocation size tested so far (64M),
based on the diminishing returns of increasing this size.
Adding batch allocation yields throughput improvement on most of the YCSB \redis{} workloads.
It also brings down the execution time of \rbench{} and \gcbench{}.
With 64M pre-allocation, batch allocation further reduces the overhead of EDMM on \gcbench{} from 41\%
in the baseline to 28\%. 

\callout{\textbf{Insight}: Batch allocation can further lower the overheads of mapping changes for memory that cannot be pre-allocated.}

\subsection{Optimization 3: Contiguous Demand Allocation ({\sf +demand<N>})}
\label{subsec:cont_demand}

\begin{figure*}[t!]
  \begin{minipage}{\textwidth}
		\includegraphics[width=.40\textwidth, height=6cm, keepaspectratio]{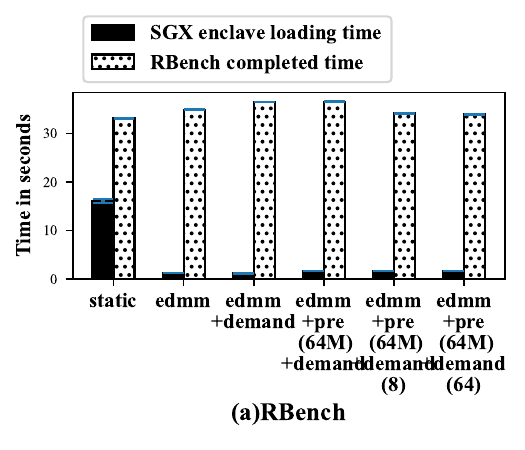}
		\hspace{5px}
		\includegraphics[width=.48\textwidth, height=6cm, keepaspectratio]{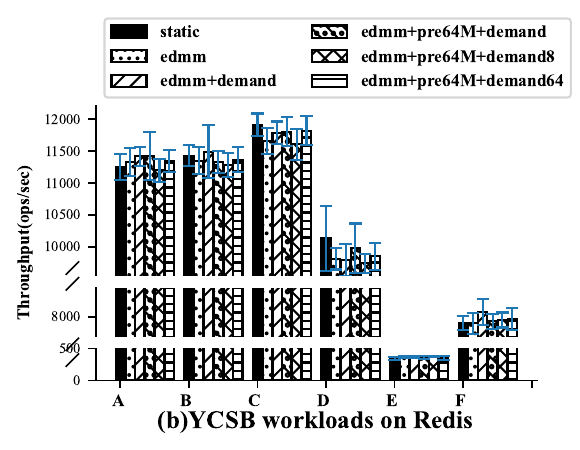}
	\end{minipage}\\
	\vspace{5px}
	\begin{minipage}{\textwidth}
		\includegraphics[width=.40\textwidth, height=6cm, keepaspectratio]{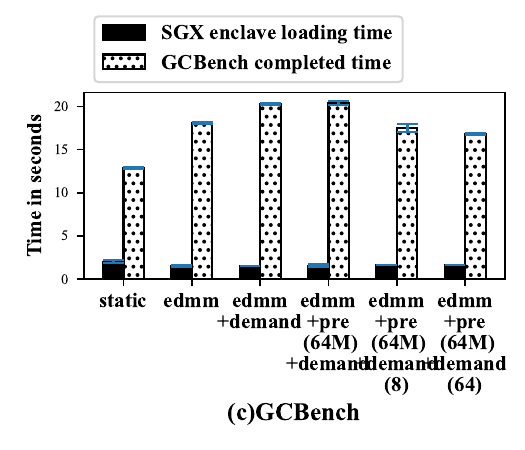}
		\hspace{5px}
		\includegraphics[width=.48\textwidth, height=6cm]{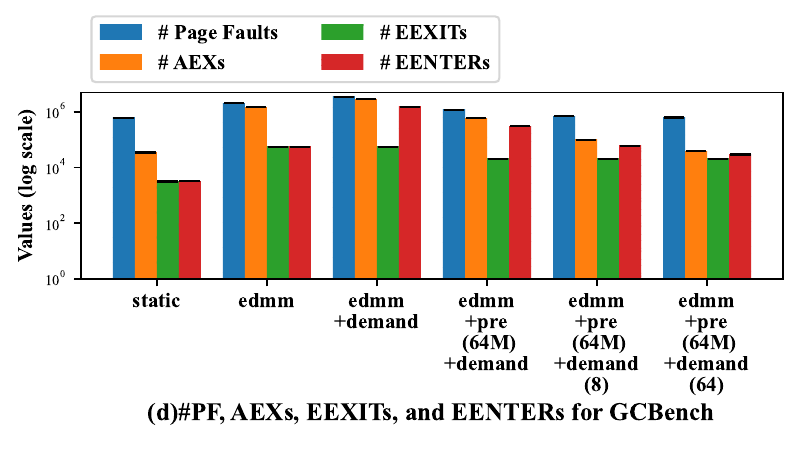}
	\end{minipage}
	\caption{({\bf Contiguous Demand Allocation}) Benchmarking of three application workloads: (a) RBench; (b) Redis;  (c) GCBench, and (d) the numbers of page faults, AEXs, EEXITs, and EENTERs during the GCBench execution.
		Each set of results is collected on the three baselines ({\sf sgx1}, {\sf edmm}, and {\sf edmm+demand}), EDMM with 64M pre-allocation ({\sf +pre(64M)}) and {\bf three different demand allocation sizes}:
	}
	\label{fig:cont_demand}
\end{figure*}

Demand allocation is a common optimization because a typical application
commonly {\tt mmap}s more virtual memory than the application accesses.
Thus, demand allocation can potentially lower enclave memory footprints.
The downside we have already illustrated is that the cost of page faults
is very high on SGX.
\fixmedp{Let me suggest a motivating experiment: how many unused pages with demand paging do you need to have before it is worth the cost of splitting a region into two mmaps?  That would basically tell you the ideal demand granularity.}
However, a high cost can potentially be amortized over more mappings.


\begin{figure}[t!]
	\includegraphics[width=.4\textwidth]{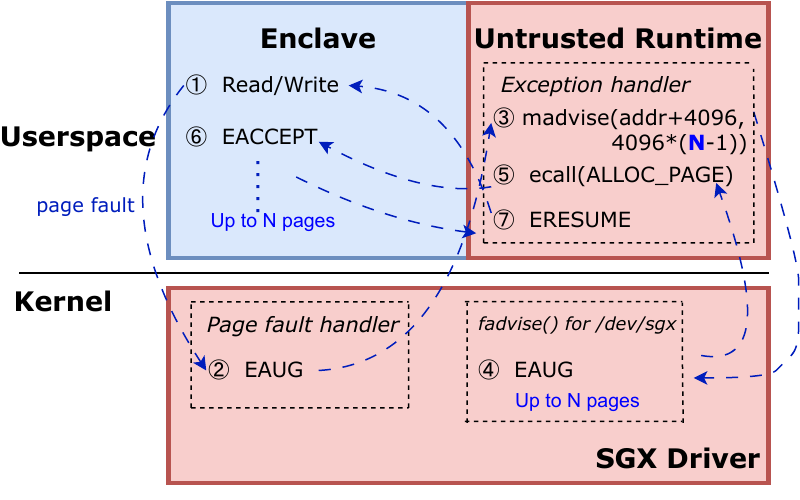}
	\caption{The optimized system flow of allocating N new enclave virtual memory mappings (N is a parameter set in the enclave's configuration) via demand allocation, involving seven context switches among the untrusted kernel, Gramine's untrusted runtime, and the enclave.  Trusted components are in blue, untrusted in pink.}\label{fig:demand-proc-multiple}
\end{figure}

{\bf Contiguous Demand Allocation} (denoted as {\sf+demand<N>} amortizes the
cost of a demand allocation fault over as many as $N$ neighboring pages.
Figure~\ref{fig:demand-proc-multiple} shows the optimized system flow to implement contiguous demand allocation.
Similar to demand allocation, the allocation process is triggered by a memory access to a virtual page that is not yet mapped by the kernel (the application has logically {\tt mmap}-ed it in Gramine).
As with baseline demand paging, the in-kernel page fault handler will call {\tt EAUG} on the faulting virtual page.
\fixmedp{Can we gain performance from doing the batch in the kernel instead of from the urts?  Seems like an experiment worth trying}
When the kernel returns to the exception handler of Gramine's untrusted runtime,
it issues the same {\tt madvise} system call as used in batched allocation to map the subsequent $N-1$ virtual pages.
\fixmedp{Shouldn't this be an aligned region, rather than the next N-1?  What if the application only requested a region smaller than N?}
Then, the untrusted runtime re-enters the enclave, which will iteratively call {\tt EACCEPT} on $N$ contiguous virtual pages starting with the faulting page.
Finally, to restore enclave state, it exits and resumes back to the original execution inside the enclave.

\vspace{0.5em}
\noindent
{\bf Evaluation.}
Figure~\ref{fig:cont_demand} shows the impact of contiguous demand allocation on our three application workloads over demand allocation, with 64M pre-allocation.
We tested with three demand allocation sizes: 1 page (baseline),
8 pages, and 64 pages.
Unsurprisingly, in the absence of memory pressure, increasing
contiguous demand allocation size lowers overheads.  In the case of
\rbench{}, demand allocating 64 pages at once effectively offsets the
cost of demand allocation to only 2\% compared to baseline EDMM
support---gaining the potential space savings of demand allocation
without the high runtime cost.
In the case of \redis{}, contiguous demand allocation
with 64 pages offsets not only the costs of demand allocation,
but EDMM in general---bringing throughput up to match static.
For \gcbench{}, contiguous demand allocation with 8 pages and 64 pages further reduces the overhead of demand allocation by 5\% and 10\%, compared to baseline EDMM.
The root cause of this improvement can be seen in Figure~\ref{fig:cont_demand}(d), in which both the number of page faults and enclave entrance and exit are reduced with contiguous demand allocation.

\callout{\textbf{Insight}: Demand paging must be at a larger granularity than one page to amortize higher mapping costs on SGX.}

\subsection{Optimization 4: Lazy Free ({\sf +lf})}

\begin{figure*}[t!]
	\begin{minipage}{\textwidth}
		\includegraphics[width=.48\textwidth]{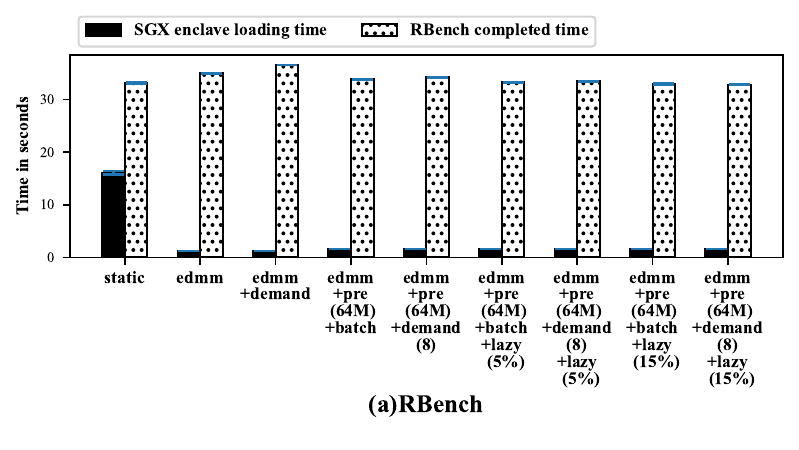}
		\hspace{5px}
		\includegraphics[width=.48\textwidth]{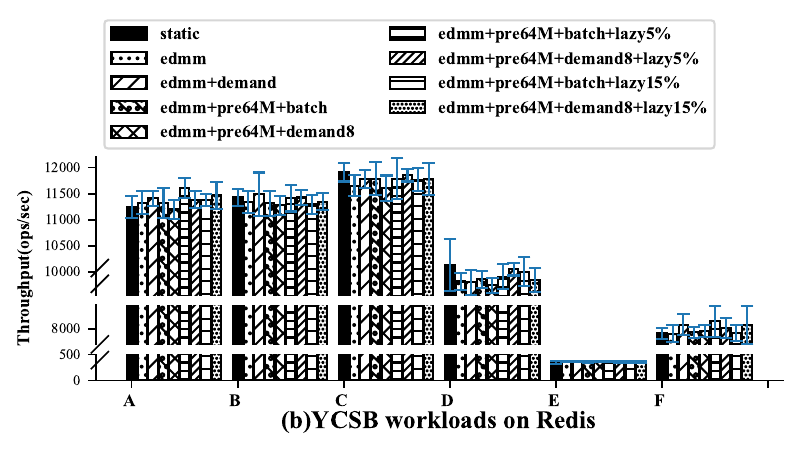}
	\end{minipage}\\
	\vspace{10px}
	\begin{minipage}{\textwidth}
		\includegraphics[width=.48\textwidth]{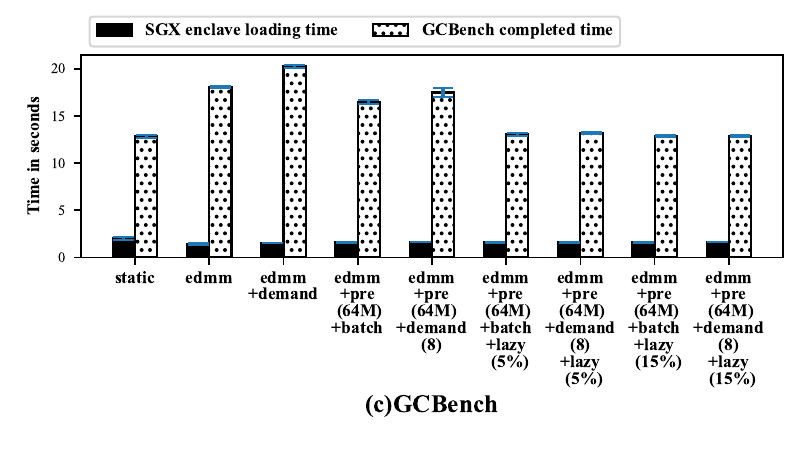}
		\hspace{5px}
		\includegraphics[width=.48\textwidth]{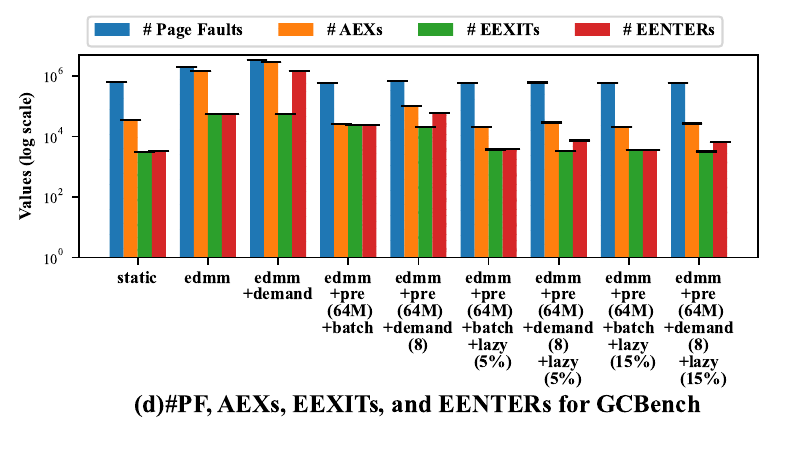}
	\end{minipage}
	\caption{({\bf Lazy Free}) Benchmarking of three application workloads: (a) RBench; (b) Redis;  (c) GCBench, and (d) the numbers of page faults, AEXs, EEXITs, and EENTERs during the GCBench execution.
		Each set of results is collected on the three baselines ({\sf sgx1}, {\sf edmm}, and {\sf edmm+demand}), EDMM with 64M pre-allocation ({\sf +pre(64M)}) and either batch ({\sf +batch}) or demand allocation ({\sf +demand}), either with or without {\bf lazy free with 15\% threshold} ({\sf +lf(15\%)}).
	}
	\label{fig:lazyfree}
\end{figure*}

Based on the observation that unmapping enclave memory is more expensive than allocating it,
the lazy free optimization caches some number of freed pages in a pool to serve subsequent
allocation requests.
We add a manifest configuration option where the user can set a maximum amount of freed memory to
hold in reserve for future allocations.
We also add bookkeeping for the page pool to track the state of free pages (allocated, unmapped, or cached).
When an application issues an {\tt munmap} to Gramine, if the cached page count is below the threshold,
these pages change state from allocated to cached. When the cached page count goes above the threshold,
pages are unmapped accordingly.

When an application issues an {\tt mmap} to Gramine, Gra\-mine first checks for a cached page region of an appropriate size;
if the allocation can be satisfied, cached pages are used.  Otherwise, new pages are demand faulted in,
as in prior subsections.

Lazy free does risk exacerbating internal fragmentation of the virtual address space over time.
However, the virtual address space of an enclave can be large and sparse.
In the case where there are enough total free physical pages, but an allocation by the OS fails,
one can simply free all of the cached pages and try again.


Another important caveat to this optimization is that it is not strictly POSIX compliant.
An application that attempts to access an {\tt munmap}-ed region should page fault.
We believe that applications deliberately faulting on an unmapped page are rare in practice.
For these rare cases, this optimization should be disabled.

\vspace{0.5em}
\noindent
{\bf Evaluation.}
Figure~\ref{fig:lazyfree} shows the impact of lazy free on our three application workloads with the optimizations we explored so far.
In particular, we test the strategy of lazy free on both batch allocation and contiguous demand allocation (8 pages at each page fault), and compare with the results in which lazy free is disabled.
We also choose two different de-allocation threshold, 5\% and 15\%, to control the {\em eagerness} of lazy free. We also tested with 64 pages at each page fault instead of 8, but did not find a significant difference.

For \gcbench{}, with a very modest threshold of 5\%, the execution time overhead drops from 28\% to a 1.8\% gain with batch allocation, and drops from 36\% to a 2.8\% gain with 8-page contiguous demand allocation.
Increasing the threshold to 15\% further improves the \gcbench{} execution time performance on par with static allocation.
For \rbench{}, increasing the threshold to 15\% even yields improved performance over static allocation in both cases with batch allocation and contiguous demand allocation.
For \redis{}, lazy free also yields improved performance on YCSB workload A and F over static allocation.

Figure~\ref{fig:lazyfree}(d) shows the reduction of enclave exits and entrances during the execution of \gcbench{} with lazy free,
which contributes to the performance improvement.


\callout{\textbf{Insight}: For applications that aggressively allocate and free virtual pages, lazy free further eliminates the remaining overheads of EDMM.}

\section{Related Work}
\label{sec:related}

The memory management of hardware enclaves has been long explored by previous works.
As one of the earliest work, Eleos~\cite{orenbach17eleos}
introduces the use of Secure User-Managed Virtual Memory (SUVM) to swap the virtual pages in and out of an enclave, without relying the untrusted SGX driver to swap the pages.
Eleos shows significant performance benefits (up to 2.3$\times$ throughput), by eliminating the cost of context switching and the subsequent cache pollution. 
A key requirement for Eleos is that the application
be compiled with an indirection mechanism for pointers, wherein the trusted runtime system
could intercept dereferencing of pointers to objects that are swapped out.
CoSMIX~\cite{cosmix} extends the idea of using a compilation pass to indirect pointers
with a software-managed oblivious RAM (ORAM), to hide page access patterns
from the untrusted OS and defend against controlled-channel attacks~\cite{Xu2015} and other
side-channel attacks~\cite{haehnel17sidechannel, van17telling, goetzfried17cache, van17sgx}.
As a follow-up work,
Autarky~\cite{autarky} explore self-paging in enclaves as a way to eliminate
controlled-channel attacks, as an extension to SGX~\cite{Xu2015}.  
They introduce a page cluster abstraction, which must be swapped as a group, and a cooperative
page management framework.

VAULT~\cite{taasori18vault} introduces
architectural changes to reduce the cost of page swapping in and out of the Enclave Page Cache (EPC). VAULT replaces Intel's SGX Integrity Tree, an in-memory data structure for authenticating the values and versions of enclave memory, with a Variable Arity Unified Tree, significantly reducing the memory accesses necessary to verify a virtual memory block.


Civet~\cite{tsai20civet}
explores enclave-aware garbage collection in the OpenJDK runtime.  Specifically, Civet introduces a three-generational
garbage collection design, which corresponds to the primary performance regimes in an enclave:
one generation fits in last-level cache (LLC), another generation fits in EPC, and the even larger, oldest generation.


\fixmedp{Tao: can you do a broader search on batched page allocation/deallocation/swapping and self-paging?}


Liu et al.~\cite{liu20preloading} show how swapping overheads in SGX can be alleviated
by preloading pages likely to be accessed in the future into the Enclave Page Cache,
either based on source analysis or observed behavior in prior execution.


\fixmedp{
TEEs + memory management
* What do non SGX TEEs do? What about trustZone?  SecureBlue? Keystone?}
\fixmees{This paper evaluates Gramine with SGX, though seemingly not EDMM: https://arxiv.org/pdf/2205.06415.pdf}

RISC-V
Keystone~\cite{keystone-enclave} uses the Linux memory allocator to assign memory to its enclaves.
Page faults caused during memory management incur high overheads, dampening the performance when stressed.
Ashman~\cite{Ashman} performs dynamic memory management on RISC-V by moving the mappings of the enclaves in order to avoid fragmentation, such that
allocated and free memory are contiguous.

Occlum~\cite{Shen2020-ja} is another library OS akin to Gramine, albeit written in Rust, which utilizes Intel Memory Protection eXtensions (MPX) to isolate memory regions.
Without EDMM, it has a better overall average performance than Gramine without EDMM.
As EDMM degrades the performance of SGX-based enclave, we show that our optimizations with Gramine-SGX reduce these overheads over baseline EDMM.
Occlum has an up-to-date EDMM implementation, and we leave a comparison between Occlum and Gramine with EDMM as future work.

Elasticlave~\cite{yu22elasticlave} is an extension to Keystone providing first-class support for sharing pages between enclaves. Unlike SGX, Elasticlave allows enclaves to selectively share memory to other enclaves, reducing overhead, as well as natively guaranteeing atomicity. 

\section{Conclusion}
In this paper, we have examined the performance impact of dynamic memory management
strategies on enclave applications, including the effect on garbage collection
and server workloads.
The results of our experiments show that, despite the reduction of enclave start time,
na\"ive implementations of dynamic memory allocation, either at user-level
mapping or at first access, cause significant slowdowns to runtime. We
have demonstrated that by optimizing the system flow of demand allocation and lazy
freeing, the runtime overhead of dynamic memory management can be significantly
reduced to being comparable to workloads under static memory allocation.

\section*{Acknowledgments}

This work was supported in part by NSF grant CNS-2244937, CNS-2148374, Intel, and the Confidential Computing Consortium (CCC).
Porter also has a significant financial interest in Fortanix and Amazon.

\bibliographystyle{ACM-Reference-Format}
\bibliography{ms}


\begin{thebibliography}{00}


\ifx \showCODEN    \undefined \def \showCODEN     #1{\unskip}     \fi
\ifx \showDOI      \undefined \def \showDOI       #1{#1}\fi
\ifx \showISBNx    \undefined \def \showISBNx     #1{\unskip}     \fi
\ifx \showISBNxiii \undefined \def \showISBNxiii  #1{\unskip}     \fi
\ifx \showISSN     \undefined \def \showISSN      #1{\unskip}     \fi
\ifx \showLCCN     \undefined \def \showLCCN      #1{\unskip}     \fi
\ifx \shownote     \undefined \def \shownote      #1{#1}          \fi
\ifx \showarticletitle \undefined \def \showarticletitle #1{#1}   \fi
\ifx \showURL      \undefined \def \showURL       {\relax}        \fi
\providecommand\bibfield[2]{#2}
\providecommand\bibinfo[2]{#2}
\providecommand\natexlab[1]{#1}
\providecommand\showeprint[2][]{arXiv:#2}

\bibitem[\protect\citeauthoryear{Alibaba}{Alibaba}{2020}]%
        {alibaba-sgx2}
\bibfield{author}{\bibinfo{person}{Alibaba}.} \bibinfo{year}{2020}\natexlab{}.
\newblock \bibinfo{title}{Alibaba Cloud Released Industry's First Trusted and
  Virtualized Instance with Support for SGX 2.0 and TPM}.
\newblock
  \bibinfo{howpublished}{\url{https://www.alibabacloud.com/blog/alibaba-cloud-released-industrys-first-trusted-and-virtualized-instance-with-support-for-sgx-2-0-and-tpm_596821}}.
    (\bibinfo{date}{October} \bibinfo{year}{2020}).
\newblock


\bibitem[\protect\citeauthoryear{Alibaba}{Alibaba}{2023}]%
        {alibaba-sgx}
\bibfield{author}{\bibinfo{person}{Alibaba}.} \bibinfo{year}{2023}\natexlab{}.
\newblock \bibinfo{title}{Alibaba Cloud, Elastic Compute Services, Instance
  Type Families, Overview}.
\newblock
  \bibinfo{howpublished}{\url{https://www.alibabacloud.com/help/doc-detail/60576.htm?spm=a2c63.p38356.b99.95.32ae1160CQKT0I}}.
    (\bibinfo{date}{August} \bibinfo{year}{2023}).
\newblock


\bibitem[\protect\citeauthoryear{{AMD}}{{AMD}}{[n. d.]}]%
        {amd-sev}
\bibfield{author}{\bibinfo{person}{{AMD}}.} \bibinfo{year}{[n. d.]}\natexlab{}.
\newblock \bibinfo{title}{{AMD Secure Encrypted Virtualization (SEV)}}.
\newblock \bibinfo{howpublished}{\url{https://developer.amd.com/sev/}}.
  (\bibinfo{year}{[n. d.]}).
\newblock


\bibitem[\protect\citeauthoryear{{AMD}}{{AMD}}{2020}]%
        {amd-sev-snp}
\bibfield{author}{\bibinfo{person}{{AMD}}.} \bibinfo{year}{2020}\natexlab{}.
\newblock \bibinfo{title}{{White PaperAMD SEV-SNP: Strengthening VM
  Isolationwith Integrity Protection and More}}.
\newblock
  \bibinfo{howpublished}{\url{https://www.amd.com/system/files/TechDocs/SEV-SNP-strengthening-vm-isolation-with-integrity-protection-and-more.pdf}}.
    (\bibinfo{year}{2020}).
\newblock


\bibitem[\protect\citeauthoryear{{ARM}}{{ARM}}{[n. d.]}]%
        {arm-secureip}
\bibfield{author}{\bibinfo{person}{{ARM}}.} \bibinfo{year}{[n. d.]}\natexlab{}.
\newblock \bibinfo{title}{{ARM Secure IP}}.
\newblock
  \bibinfo{howpublished}{\url{https://developer.arm.com/ip-products/security-ip}}.
    (\bibinfo{year}{[n. d.]}).
\newblock


\bibitem[\protect\citeauthoryear{Arnautov, Trach, Gregor, Knauth, Martin,
  Priebe, Lind, Muthukumaran, O'Keeffe, Stillwell, Goltzsche, Eyers, Kapitza,
  Pietzuch, and Fetzer}{Arnautov et~al\mbox{.}}{2016}]%
        {Arnautov2016}
\bibfield{author}{\bibinfo{person}{Sergei Arnautov}, \bibinfo{person}{Bohdan
  Trach}, \bibinfo{person}{Franz Gregor}, \bibinfo{person}{Thomas Knauth},
  \bibinfo{person}{Andre Martin}, \bibinfo{person}{Christian Priebe},
  \bibinfo{person}{Joshua Lind}, \bibinfo{person}{Divya Muthukumaran},
  \bibinfo{person}{Dan O'Keeffe}, \bibinfo{person}{Mark~L. Stillwell},
  \bibinfo{person}{David Goltzsche}, \bibinfo{person}{David Eyers},
  \bibinfo{person}{R\"{u}diger Kapitza}, \bibinfo{person}{Peter Pietzuch},
  {and} \bibinfo{person}{Christof Fetzer}.} \bibinfo{year}{2016}\natexlab{}.
\newblock \showarticletitle{SCONE: Secure Linux Containers with Intel SGX}. In
  \bibinfo{booktitle}{{\em Proceedings of the 12th USENIX Conference on
  Operating Systems Design and Implementation}} {\em
  (\bibinfo{series}{OSDI'16})}. \bibinfo{publisher}{USENIX Association},
  \bibinfo{address}{Berkeley, CA, USA}, \bibinfo{pages}{689--703}.
\newblock
\showISBNx{978-1-931971-33-1}
\showURL{%
\url{http://dl.acm.org/citation.cfm?id=3026877.3026930}}


\bibitem[\protect\citeauthoryear{Baumann, Peinado, and Hunt}{Baumann
  et~al\mbox{.}}{2014}]%
        {Baumann2014}
\bibfield{author}{\bibinfo{person}{Andrew Baumann}, \bibinfo{person}{Marcus
  Peinado}, {and} \bibinfo{person}{Galen Hunt}.}
  \bibinfo{year}{2014}\natexlab{}.
\newblock \showarticletitle{Shielding Applications from an Untrusted Cloud with
  Haven}. In \bibinfo{booktitle}{{\em 11th {USENIX} Symposium on Operating
  Systems Design and Implementation ({OSDI} 14)}}. \bibinfo{publisher}{{USENIX}
  Association}, \bibinfo{address}{Broomfield, CO}, \bibinfo{pages}{267--283}.
\newblock
\showISBNx{978-1-931971-16-4}
\showURL{%
\url{https://www.usenix.org/conference/osdi14/technical-sessions/presentation/baumann}}


\bibitem[\protect\citeauthoryear{Brenner, Wulf, Goltzsche, Weichbrodt, Lorenz,
  Fetzer, Pietzuch, and Kapitza}{Brenner et~al\mbox{.}}{2016}]%
        {SecureKeeper}
\bibfield{author}{\bibinfo{person}{Stefan Brenner}, \bibinfo{person}{Colin
  Wulf}, \bibinfo{person}{David Goltzsche}, \bibinfo{person}{Nico Weichbrodt},
  \bibinfo{person}{Matthias Lorenz}, \bibinfo{person}{Christof Fetzer},
  \bibinfo{person}{Peter Pietzuch}, {and} \bibinfo{person}{R\"{u}diger
  Kapitza}.} \bibinfo{year}{2016}\natexlab{}.
\newblock \showarticletitle{SecureKeeper: Confidential ZooKeeper Using Intel
  SGX}. In \bibinfo{booktitle}{{\em Proceedings of the 17th International
  Middleware Conference}} {\em (\bibinfo{series}{Middleware '16})}.
  \bibinfo{publisher}{Association for Computing Machinery},
  \bibinfo{address}{New York, NY, USA}, Article \bibinfo{articleno}{14},
  \bibinfo{numpages}{13}~pages.
\newblock
\showDOI{%
\url{https://doi.org/10.1145/2988336.2988350}}


\bibitem[\protect\citeauthoryear{Eskandarian and Zaharia}{Eskandarian and
  Zaharia}{2019}]%
        {ObliDB}
\bibfield{author}{\bibinfo{person}{Saba Eskandarian} {and}
  \bibinfo{person}{Matei Zaharia}.} \bibinfo{year}{2019}\natexlab{}.
\newblock \showarticletitle{ObliDB: Oblivious Query Processing for Secure
  Databases}.
\newblock \bibinfo{journal}{{\em Proc. VLDB Endow.\/}} \bibinfo{volume}{13},
  \bibinfo{number}{2} (\bibinfo{date}{Oct.} \bibinfo{year}{2019}),
  \bibinfo{pages}{169–183}.
\newblock
\showISSN{2150-8097}
\showDOI{%
\url{https://doi.org/10.14778/3364324.3364331}}


\bibitem[\protect\citeauthoryear{Fisch, Vinayagamurthy, Boneh, and
  Gorbunov}{Fisch et~al\mbox{.}}{2017}]%
        {Firch17}
\bibfield{author}{\bibinfo{person}{Ben Fisch}, \bibinfo{person}{Dhinakaran
  Vinayagamurthy}, \bibinfo{person}{Dan Boneh}, {and} \bibinfo{person}{Sergey
  Gorbunov}.} \bibinfo{year}{2017}\natexlab{}.
\newblock \showarticletitle{IRON: Functional Encryption Using Intel SGX}. In
  \bibinfo{booktitle}{{\em Proceedings of the 2017 ACM SIGSAC Conference on
  Computer and Communications Security}} {\em (\bibinfo{series}{CCS '17})}.
  \bibinfo{publisher}{Association for Computing Machinery},
  \bibinfo{address}{New York, NY, USA}, \bibinfo{pages}{765–782}.
\newblock
\showISBNx{9781450349468}
\showDOI{%
\url{https://doi.org/10.1145/3133956.3134106}}


\bibitem[\protect\citeauthoryear{G\"{o}tzfried, Eckert, Schinzel, and
  M\"{u}ller}{G\"{o}tzfried et~al\mbox{.}}{2017}]%
        {goetzfried17cache}
\bibfield{author}{\bibinfo{person}{Johannes G\"{o}tzfried},
  \bibinfo{person}{Moritz Eckert}, \bibinfo{person}{Sebastian Schinzel}, {and}
  \bibinfo{person}{Tilo M\"{u}ller}.} \bibinfo{year}{2017}\natexlab{}.
\newblock \showarticletitle{Cache Attacks on Intel SGX}. In
  \bibinfo{booktitle}{{\em Proceedings of the 10th European Workshop on Systems
  Security}} {\em (\bibinfo{series}{EuroSec'17})}.
  \bibinfo{publisher}{Association for Computing Machinery},
  \bibinfo{address}{New York, NY, USA}, Article \bibinfo{articleno}{2},
  \bibinfo{numpages}{6}~pages.
\newblock
\showISBNx{9781450349352}
\showDOI{%
\url{https://doi.org/10.1145/3065913.3065915}}


\bibitem[\protect\citeauthoryear{H{\"a}hnel, Cui, and Peinado}{H{\"a}hnel
  et~al\mbox{.}}{2017}]%
        {haehnel17sidechannel}
\bibfield{author}{\bibinfo{person}{Marcus H{\"a}hnel}, \bibinfo{person}{Weidong
  Cui}, {and} \bibinfo{person}{Marcus Peinado}.}
  \bibinfo{year}{2017}\natexlab{}.
\newblock \showarticletitle{High-Resolution Side Channels for Untrusted
  Operating Systems}. In \bibinfo{booktitle}{{\em 2017 {USENIX} Annual
  Technical Conference ({USENIX} {ATC} 17)}}. \bibinfo{publisher}{{USENIX}
  Association}, \bibinfo{address}{Santa Clara, CA}, \bibinfo{pages}{299--312}.
\newblock
\showISBNx{978-1-931971-38-6}
\showURL{%
\url{https://www.usenix.org/conference/atc17/technical-sessions/presentation/hahnel}}


\bibitem[\protect\citeauthoryear{IBM}{IBM}{2020}]%
        {ibm-datashield}
\bibfield{author}{\bibinfo{person}{IBM}.} \bibinfo{year}{2020}\natexlab{}.
\newblock \bibinfo{title}{IBM Cloud Data Shield Now Generally Available}.
\newblock
  \bibinfo{howpublished}{\url{https://www.ibm.com/blog/announcement/ibm-cloud-data-shield-now-generally-available/}}.
    (\bibinfo{date}{April} \bibinfo{year}{2020}).
\newblock


\bibitem[\protect\citeauthoryear{IBM}{IBM}{2023}]%
        {ibm-sgx}
\bibfield{author}{\bibinfo{person}{IBM}.} \bibinfo{year}{2023}\natexlab{}.
\newblock \bibinfo{title}{Provisioning a bare metal server with Intel®
  Software Guard Extension architecture}.
\newblock
  \bibinfo{howpublished}{\url{https://cloud.ibm.com/docs/bare-metal?topic=bare-metal-bm-server-provision-sgx}}.
    (\bibinfo{date}{January} \bibinfo{year}{2023}).
\newblock


\bibitem[\protect\citeauthoryear{{Intel}}{{Intel}}{2021}]%
        {tme-whitepaper}
\bibfield{author}{\bibinfo{person}{{Intel}}.} \bibinfo{year}{2021}\natexlab{}.
\newblock \showarticletitle{Intel\&Reg; Hardware Shield--Intel\&Reg; Total
  Memory Encryption}.
\newblock
  \bibinfo{howpublished}{\url{https://www.intel.com/content/dam/www/central-libraries/us/en/documents/white-paper-intel-tme.pdf}}.
\newblock  (\bibinfo{year}{2021}).
\newblock


\bibitem[\protect\citeauthoryear{Intel}{Intel}{2022a}]%
        {aex-notify-wp}
\bibfield{author}{\bibinfo{person}{Intel}.} \bibinfo{year}{2022}\natexlab{a}.
\newblock \bibinfo{title}{Asynchronous Enclave Exit Notify and the EDECCSSA
  User Leaf Function}.
\newblock
  \bibinfo{howpublished}{\url{https://cdrdv2.intel.com/v1/dl/getContent/736463?explicitVersion=true}}.
    (\bibinfo{year}{2022}).
\newblock


\bibitem[\protect\citeauthoryear{Intel}{Intel}{2022b}]%
        {tdx}
\bibfield{author}{\bibinfo{person}{Intel}.} \bibinfo{year}{2022}\natexlab{b}.
\newblock \bibinfo{title}{Intel Trust Domain Extensions}.
\newblock
  \bibinfo{howpublished}{\url{https://cdrdv2.intel.com/v1/dl/getContent/690419}}.
    (\bibinfo{year}{2022}).
\newblock


\bibitem[\protect\citeauthoryear{Kim, Kim, Rhee, Yu, Chen, Tian, and Lee}{Kim
  et~al\mbox{.}}{2020}]%
        {Vessels}
\bibfield{author}{\bibinfo{person}{Kyungtae Kim}, \bibinfo{person}{Chung~Hwan
  Kim}, \bibinfo{person}{Junghwan~"John" Rhee}, \bibinfo{person}{Xiao Yu},
  \bibinfo{person}{Haifeng Chen}, \bibinfo{person}{Dave~(Jing) Tian}, {and}
  \bibinfo{person}{Byoungyoung Lee}.} \bibinfo{year}{2020}\natexlab{}.
\newblock \showarticletitle{Vessels: Efficient and Scalable Deep Learning
  Prediction on Trusted Processors}. In \bibinfo{booktitle}{{\em Proceedings of
  the 11th ACM Symposium on Cloud Computing}} {\em (\bibinfo{series}{SoCC
  '20})}. \bibinfo{publisher}{Association for Computing Machinery},
  \bibinfo{address}{New York, NY, USA}, \bibinfo{pages}{462–476}.
\newblock
\showISBNx{9781450381376}
\showDOI{%
\url{https://doi.org/10.1145/3419111.3421282}}


\bibitem[\protect\citeauthoryear{Kim, Han, Ha, Kim, and Han}{Kim
  et~al\mbox{.}}{2017}]%
        {Kim17}
\bibfield{author}{\bibinfo{person}{Seongmin Kim}, \bibinfo{person}{Juhyeng
  Han}, \bibinfo{person}{Jaehyeong Ha}, \bibinfo{person}{Taesoo Kim}, {and}
  \bibinfo{person}{Dongsu Han}.} \bibinfo{year}{2017}\natexlab{}.
\newblock \showarticletitle{Enhancing Security and Privacy of
  Tor{\textquoteright}s Ecosystem by Using Trusted Execution Environments}. In
  \bibinfo{booktitle}{{\em 14th {USENIX} Symposium on Networked Systems Design
  and Implementation ({NSDI} 17)}}. \bibinfo{publisher}{{USENIX} Association},
  \bibinfo{address}{Boston, MA}, \bibinfo{pages}{145--161}.
\newblock
\showISBNx{978-1-931971-37-9}
\showURL{%
\url{https://www.usenix.org/conference/nsdi17/technical-sessions/presentation/kim-seongmin}}


\bibitem[\protect\citeauthoryear{Kim, Park, Woo, Jeon, and Huh}{Kim
  et~al\mbox{.}}{2019}]%
        {ShieldStore}
\bibfield{author}{\bibinfo{person}{Taehoon Kim}, \bibinfo{person}{Joongun
  Park}, \bibinfo{person}{Jaewook Woo}, \bibinfo{person}{Seungheun Jeon}, {and}
  \bibinfo{person}{Jaehyuk Huh}.} \bibinfo{year}{2019}\natexlab{}.
\newblock \showarticletitle{ShieldStore: Shielded In-Memory Key-Value Storage
  with SGX}. In \bibinfo{booktitle}{{\em Proceedings of the Fourteenth EuroSys
  Conference 2019}} {\em (\bibinfo{series}{EuroSys '19})}.
  \bibinfo{publisher}{Association for Computing Machinery},
  \bibinfo{address}{New York, NY, USA}, Article \bibinfo{articleno}{14},
  \bibinfo{numpages}{15}~pages.
\newblock
\showISBNx{9781450362818}
\showDOI{%
\url{https://doi.org/10.1145/3302424.3303951}}


\bibitem[\protect\citeauthoryear{K\"{u}\c{c}\"{u}k, Paverd, Martin, Asokan,
  Simpson, and Ankele}{K\"{u}\c{c}\"{u}k et~al\mbox{.}}{2016}]%
        {Kuccuk16}
\bibfield{author}{\bibinfo{person}{Kubilay~Ahmet K\"{u}\c{c}\"{u}k},
  \bibinfo{person}{Andrew Paverd}, \bibinfo{person}{Andrew Martin},
  \bibinfo{person}{N. Asokan}, \bibinfo{person}{Andrew Simpson}, {and}
  \bibinfo{person}{Robin Ankele}.} \bibinfo{year}{2016}\natexlab{}.
\newblock \showarticletitle{Exploring the Use of Intel SGX for Secure
  Many-Party Applications}. In \bibinfo{booktitle}{{\em Proceedings of the 1st
  Workshop on System Software for Trusted Execution}} {\em
  (\bibinfo{series}{SysTEX '16})}. \bibinfo{publisher}{Association for
  Computing Machinery}, \bibinfo{address}{New York, NY, USA}, Article
  \bibinfo{articleno}{5}, \bibinfo{numpages}{6}~pages.
\newblock
\showISBNx{9781450346702}
\showDOI{%
\url{https://doi.org/10.1145/3007788.3007793}}


\bibitem[\protect\citeauthoryear{Lee, Kohlbrenner, Shinde, Asanovi\'{c}, and
  Song}{Lee et~al\mbox{.}}{2020}]%
        {keystone-enclave}
\bibfield{author}{\bibinfo{person}{Dayeol Lee}, \bibinfo{person}{David
  Kohlbrenner}, \bibinfo{person}{Shweta Shinde}, \bibinfo{person}{Krste
  Asanovi\'{c}}, {and} \bibinfo{person}{Dawn Song}.}
  \bibinfo{year}{2020}\natexlab{}.
\newblock \showarticletitle{Keystone: An Open Framework for Architecting
  Trusted Execution Environments}. In \bibinfo{booktitle}{{\em Proceedings of
  the Fifteenth European Conference on Computer Systems}} {\em
  (\bibinfo{series}{EuroSys '20})}. \bibinfo{publisher}{Association for
  Computing Machinery}, \bibinfo{address}{New York, NY, USA}, Article
  \bibinfo{articleno}{38}, \bibinfo{numpages}{16}~pages.
\newblock
\showISBNx{9781450368827}
\showDOI{%
\url{https://doi.org/10.1145/3342195.3387532}}


\bibitem[\protect\citeauthoryear{Li, Huang, Ren, Lu, Ning, Cui, and Zhang}{Li
  et~al\mbox{.}}{2022}]%
        {Ashman}
\bibfield{author}{\bibinfo{person}{Haonan Li}, \bibinfo{person}{Weijie Huang},
  \bibinfo{person}{Mingde Ren}, \bibinfo{person}{Hongyi Lu},
  \bibinfo{person}{Zhenyu Ning}, \bibinfo{person}{Heming Cui}, {and}
  \bibinfo{person}{Fengwei Zhang}.} \bibinfo{year}{2022}\natexlab{}.
\newblock \showarticletitle{A Novel Memory Management for RISC-V Enclaves}. In
  \bibinfo{booktitle}{{\em Proceedings of the 10th International Workshop on
  Hardware and Architectural Support for Security and Privacy}} {\em
  (\bibinfo{series}{HASP '21})}. \bibinfo{publisher}{Association for Computing
  Machinery}, \bibinfo{address}{New York, NY, USA}, Article
  \bibinfo{articleno}{3}, \bibinfo{numpages}{9}~pages.
\newblock
\showISBNx{9781450396141}
\showDOI{%
\url{https://doi.org/10.1145/3505253.3505257}}


\bibitem[\protect\citeauthoryear{Liu, Wang, Wang, Gong, Zhao, and Yew}{Liu
  et~al\mbox{.}}{2020}]%
        {liu20preloading}
\bibfield{author}{\bibinfo{person}{Ximing Liu}, \bibinfo{person}{Wenwen Wang},
  \bibinfo{person}{Lizhi Wang}, \bibinfo{person}{Xiaoli Gong},
  \bibinfo{person}{Ziyi Zhao}, {and} \bibinfo{person}{Pen-Chung Yew}.}
  \bibinfo{year}{2020}\natexlab{}.
\newblock \showarticletitle{Regaining Lost Seconds: Efficient Page Preloading
  for SGX Enclaves}. In \bibinfo{booktitle}{{\em Proceedings of the 21st
  International Middleware Conference}} {\em (\bibinfo{series}{Middleware
  '20})}. \bibinfo{publisher}{Association for Computing Machinery},
  \bibinfo{address}{New York, NY, USA}, \bibinfo{pages}{326–340}.
\newblock
\showISBNx{9781450381536}
\showDOI{%
\url{https://doi.org/10.1145/3423211.3425673}}


\bibitem[\protect\citeauthoryear{McKeen, Alexandrovich, Anati, Caspi, Johnson,
  Leslie-Hurd, and Rozas}{McKeen et~al\mbox{.}}{2016}]%
        {McKeen2016}
\bibfield{author}{\bibinfo{person}{Frank McKeen}, \bibinfo{person}{Ilya
  Alexandrovich}, \bibinfo{person}{Ittai Anati}, \bibinfo{person}{Dror Caspi},
  \bibinfo{person}{Simon Johnson}, \bibinfo{person}{Rebekah Leslie-Hurd}, {and}
  \bibinfo{person}{Carlos Rozas}.} \bibinfo{year}{2016}\natexlab{}.
\newblock \showarticletitle{Intel\&Reg; Software Guard Extensions (Intel\&Reg;
  SGX) Support for Dynamic Memory Management Inside an Enclave}. In
  \bibinfo{booktitle}{{\em Proceedings of the Hardware and Architectural
  Support for Security and Privacy 2016}} {\em (\bibinfo{series}{HASP 2016})}.
  \bibinfo{publisher}{ACM}, \bibinfo{address}{New York, NY, USA}, Article
  \bibinfo{articleno}{10}, \bibinfo{numpages}{9}~pages.
\newblock
\showISBNx{978-1-4503-4769-3}
\showDOI{%
\url{https://doi.org/10.1145/2948618.2954331}}


\bibitem[\protect\citeauthoryear{McKeen, Alexandrovich, Berenzon, Rozas, Shafi,
  Shanbhogue, and Savagaonkar}{McKeen et~al\mbox{.}}{2013}]%
        {intelsgx}
\bibfield{author}{\bibinfo{person}{Frank McKeen}, \bibinfo{person}{Ilya
  Alexandrovich}, \bibinfo{person}{Alex Berenzon}, \bibinfo{person}{Carlos~V.
  Rozas}, \bibinfo{person}{Hisham Shafi}, \bibinfo{person}{Vedvyas Shanbhogue},
  {and} \bibinfo{person}{Uday~R. Savagaonkar}.}
  \bibinfo{year}{2013}\natexlab{}.
\newblock \showarticletitle{Innovative Instructions and Software Model for
  Isolated Execution}. In \bibinfo{booktitle}{{\em Proceedings of the 2nd
  International Workshop on Hardware and Architectural Support for Security and
  Privacy ({{HASP} 13})}}. \bibinfo{publisher}{ACM}.
\newblock
\showURL{%
\url{http://doi.acm.org/10.1145/2487726.2488368}}


\bibitem[\protect\citeauthoryear{Microsoft}{Microsoft}{2023}]%
        {microsoft-sgx-vms}
\bibfield{author}{\bibinfo{person}{Microsoft}.}
  \bibinfo{year}{2023}\natexlab{}.
\newblock \bibinfo{title}{DCsv3 and DCdsv3-series}.
\newblock
  \bibinfo{howpublished}{\url{https://learn.microsoft.com/en-us/azure/virtual-machines/dcv3-series}}.
    (\bibinfo{date}{January} \bibinfo{year}{2023}).
\newblock


\bibitem[\protect\citeauthoryear{Mishra, Poddar, Chen, Chiesa, and Popa}{Mishra
  et~al\mbox{.}}{2018}]%
        {Oblix}
\bibfield{author}{\bibinfo{person}{Pratyush Mishra}, \bibinfo{person}{Rishabh
  Poddar}, \bibinfo{person}{Jerry Chen}, \bibinfo{person}{Alessandro Chiesa},
  {and} \bibinfo{person}{Raluca Popa}.} \bibinfo{year}{2018}\natexlab{}.
\newblock \showarticletitle{Oblix: An Efficient Oblivious Search Index}.
  \bibinfo{pages}{279--296}.
\newblock
\showDOI{%
\url{https://doi.org/10.1109/SP.2018.00045}}


\bibitem[\protect\citeauthoryear{Orenbach, Baumann, and Silberstein}{Orenbach
  et~al\mbox{.}}{2020}]%
        {autarky}
\bibfield{author}{\bibinfo{person}{Meni Orenbach}, \bibinfo{person}{Andrew
  Baumann}, {and} \bibinfo{person}{Mark Silberstein}.}
  \bibinfo{year}{2020}\natexlab{}.
\newblock \showarticletitle{Autarky: Closing Controlled Channels with
  Self-Paging Enclaves}. In \bibinfo{booktitle}{{\em Proceedings of the
  Fifteenth European Conference on Computer Systems}} {\em
  (\bibinfo{series}{EuroSys '20})}. \bibinfo{publisher}{Association for
  Computing Machinery}, \bibinfo{address}{New York, NY, USA}, Article
  \bibinfo{articleno}{7}, \bibinfo{numpages}{16}~pages.
\newblock
\showISBNx{9781450368827}
\showDOI{%
\url{https://doi.org/10.1145/3342195.3387541}}


\bibitem[\protect\citeauthoryear{Orenbach, Lifshits, Minkin, and
  Silberstein}{Orenbach et~al\mbox{.}}{2017}]%
        {orenbach17eleos}
\bibfield{author}{\bibinfo{person}{Meni Orenbach}, \bibinfo{person}{Pavel
  Lifshits}, \bibinfo{person}{Marina Minkin}, {and} \bibinfo{person}{Mark
  Silberstein}.} \bibinfo{year}{2017}\natexlab{}.
\newblock \showarticletitle{Eleos: ExitLess OS Services for SGX Enclaves}. In
  \bibinfo{booktitle}{{\em Proceedings of the Twelfth European Conference on
  Computer Systems ({EuroSys} 17)}}. \bibinfo{pages}{238--253}.
\newblock


\bibitem[\protect\citeauthoryear{Orenbach, Michalevsky, Fetzer, and
  Silberstein}{Orenbach et~al\mbox{.}}{2019}]%
        {cosmix}
\bibfield{author}{\bibinfo{person}{Meni Orenbach}, \bibinfo{person}{Yan
  Michalevsky}, \bibinfo{person}{Christof Fetzer}, {and} \bibinfo{person}{Mark
  Silberstein}.} \bibinfo{year}{2019}\natexlab{}.
\newblock \showarticletitle{CoSMIX: A Compiler-based System for Secure Memory
  Instrumentation and Execution in Enclaves}. In \bibinfo{booktitle}{{\em 2019
  {USENIX} Annual Technical Conference ({USENIX} {ATC} 19)}}.
  \bibinfo{publisher}{{USENIX} Association}, \bibinfo{address}{Renton, WA},
  \bibinfo{pages}{555--570}.
\newblock
\showISBNx{978-1-939133-03-8}
\showURL{%
\url{https://www.usenix.org/conference/atc19/presentation/orenbach}}


\bibitem[\protect\citeauthoryear{Pedroni, Boehm, Ellis, and Kovac}{Pedroni
  et~al\mbox{.}}{2018}]%
        {gcbench}
\bibfield{author}{\bibinfo{person}{Samuele Pedroni}, \bibinfo{person}{Hans
  Boehm}, \bibinfo{person}{John Ellis}, {and} \bibinfo{person}{Pete Kovac}.}
  \bibinfo{year}{2018}\natexlab{}.
\newblock \bibinfo{title}{GCBench}.
\newblock
  \bibinfo{howpublished}{\url{https://github.com/mozillazg/pypy/blob/40795dcad7e1b0be53d2f95a94f0278086d2d448/rpython/translator/goal/gcbench.py}}.
    (\bibinfo{year}{2018}).
\newblock


\bibitem[\protect\citeauthoryear{Pires, Pasin, Felber, and Fetzer}{Pires
  et~al\mbox{.}}{2016}]%
        {Pires16}
\bibfield{author}{\bibinfo{person}{Rafael Pires}, \bibinfo{person}{Marcelo
  Pasin}, \bibinfo{person}{Pascal Felber}, {and} \bibinfo{person}{Christof
  Fetzer}.} \bibinfo{year}{2016}\natexlab{}.
\newblock \showarticletitle{Secure Content-Based Routing Using Intel Software
  Guard Extensions}. In \bibinfo{booktitle}{{\em Proceedings of the 17th
  International Middleware Conference}} {\em (\bibinfo{series}{Middleware
  '16})}. \bibinfo{publisher}{Association for Computing Machinery},
  \bibinfo{address}{New York, NY, USA}, Article \bibinfo{articleno}{10},
  \bibinfo{numpages}{10}~pages.
\newblock
\showISBNx{9781450343008}
\showDOI{%
\url{https://doi.org/10.1145/2988336.2988346}}


\bibitem[\protect\citeauthoryear{Poddar, Lan, Popa, and Ratnasamy}{Poddar
  et~al\mbox{.}}{2018}]%
        {Safebricks}
\bibfield{author}{\bibinfo{person}{Rishabh Poddar}, \bibinfo{person}{Chang
  Lan}, \bibinfo{person}{Raluca~Ada Popa}, {and} \bibinfo{person}{Sylvia
  Ratnasamy}.} \bibinfo{year}{2018}\natexlab{}.
\newblock \showarticletitle{Safebricks: Shielding Network Functions in the
  Cloud}. In \bibinfo{booktitle}{{\em Proceedings of the 15th USENIX Conference
  on Networked Systems Design and Implementation}} {\em
  (\bibinfo{series}{NSDI'18})}. \bibinfo{publisher}{USENIX Association},
  \bibinfo{address}{USA}, \bibinfo{pages}{201–216}.
\newblock
\showISBNx{9781931971430}


\bibitem[\protect\citeauthoryear{Priebe, Vaswani, and Costa}{Priebe
  et~al\mbox{.}}{2018}]%
        {priebe2018enclavedb}
\bibfield{author}{\bibinfo{person}{Christian Priebe}, \bibinfo{person}{Kapil
  Vaswani}, {and} \bibinfo{person}{Manuel Costa}.}
  \bibinfo{year}{2018}\natexlab{}.
\newblock \showarticletitle{EnclaveDB – A Secure Database using SGX}. In
  \bibinfo{booktitle}{{\em \oakland{}}}. \bibinfo{publisher}{IEEE}.
\newblock
\showURL{%
\url{https://www.microsoft.com/en-us/research/publication/enclavedb-a-secure-database-using-sgx/}}


\bibitem[\protect\citeauthoryear{{Redis}}{{Redis}}{2023}]%
        {redis}
\bibfield{author}{\bibinfo{person}{{Redis}}.} \bibinfo{year}{2023}\natexlab{}.
\newblock \bibinfo{title}{Redis benchmark}.
\newblock
  \bibinfo{howpublished}{\url{https://redis.io/docs/management/optimization/benchmarks/}}.
    (\bibinfo{year}{2023}).
\newblock


\bibitem[\protect\citeauthoryear{Schuster, Costa, Fournet, Gkantsidis, Peinado,
  Mainar-Ruiz, and Russinovich}{Schuster et~al\mbox{.}}{2015}]%
        {vc3}
\bibfield{author}{\bibinfo{person}{Felix Schuster}, \bibinfo{person}{Manuel
  Costa}, \bibinfo{person}{C{\'e}dric Fournet}, \bibinfo{person}{Christos
  Gkantsidis}, \bibinfo{person}{Marcus Peinado}, \bibinfo{person}{Gloria
  Mainar-Ruiz}, {and} \bibinfo{person}{Mark Russinovich}.}
  \bibinfo{year}{2015}\natexlab{}.
\newblock \showarticletitle{{VC3}: Trustworthy data analytics in the cloud
  using {SGX}}. In \bibinfo{booktitle}{{\em 2015 {IEEE} Symposium on Security
  and Privacy (S\&P 15)}}. IEEE, \bibinfo{pages}{38--54}.
\newblock


\bibitem[\protect\citeauthoryear{Shen, Tian, Chen, Chen, Wang, Xu, Xia, and
  Yan}{Shen et~al\mbox{.}}{2020}]%
        {Shen2020-ja}
\bibfield{author}{\bibinfo{person}{Youren Shen}, \bibinfo{person}{Hongliang
  Tian}, \bibinfo{person}{Yu Chen}, \bibinfo{person}{Kang Chen},
  \bibinfo{person}{Runji Wang}, \bibinfo{person}{Yi Xu}, \bibinfo{person}{Yubin
  Xia}, {and} \bibinfo{person}{Shoumeng Yan}.} \bibinfo{year}{2020}\natexlab{}.
\newblock \showarticletitle{Occlum: Secure and efficient multitasking inside a
  single enclave of Intel {SGX}}. In \bibinfo{booktitle}{{\em Proceedings of
  the {Twenty-Fifth} International Conference on Architectural Support for
  Programming Languages and Operating Systems}}. \bibinfo{publisher}{ACM},
  \bibinfo{address}{New York, NY, USA}.
\newblock


\bibitem[\protect\citeauthoryear{Taassori, Shafiee, and
  Balasubramonian}{Taassori et~al\mbox{.}}{2018}]%
        {taasori18vault}
\bibfield{author}{\bibinfo{person}{Meysam Taassori}, \bibinfo{person}{Ali
  Shafiee}, {and} \bibinfo{person}{Rajeev Balasubramonian}.}
  \bibinfo{year}{2018}\natexlab{}.
\newblock \showarticletitle{VAULT: Reducing Paging Overheads in SGX with
  Efficient Integrity Verification Structures}. In \bibinfo{booktitle}{{\em
  Proceedings of the Twenty-Third International Conference on Architectural
  Support for Programming Languages and Operating Systems}} {\em
  (\bibinfo{series}{ASPLOS '18})}. \bibinfo{publisher}{Association for
  Computing Machinery}, \bibinfo{address}{New York, NY, USA},
  \bibinfo{pages}{665–678}.
\newblock
\showISBNx{9781450349116}
\showDOI{%
\url{https://doi.org/10.1145/3173162.3177155}}


\bibitem[\protect\citeauthoryear{Tsai, Arora, Bandi, Jain, Jannen, John,
  Kalodner, Kulkarni, Oliveira, and Porter}{Tsai et~al\mbox{.}}{2014}]%
        {tsai14graphene}
\bibfield{author}{\bibinfo{person}{Chia-Che Tsai},
  \bibinfo{person}{Kumar~Saurabh Arora}, \bibinfo{person}{Nehal Bandi},
  \bibinfo{person}{Bhushan Jain}, \bibinfo{person}{William Jannen},
  \bibinfo{person}{Jitin John}, \bibinfo{person}{Harry~A. Kalodner},
  \bibinfo{person}{Vrushali Kulkarni}, \bibinfo{person}{Daniela Oliveira},
  {and} \bibinfo{person}{Donald~E. Porter}.} \bibinfo{year}{2014}\natexlab{}.
\newblock \showarticletitle{{Cooperation and Security Isolation of Library OSes
  for Multi-Process Applications}}. In \bibinfo{booktitle}{{\em \eurosys{}}}.
\newblock


\bibitem[\protect\citeauthoryear{Tsai, Porter, and Vij}{Tsai
  et~al\mbox{.}}{2017}]%
        {Tsai2017}
\bibfield{author}{\bibinfo{person}{Chia-Che Tsai}, \bibinfo{person}{Donald~E.
  Porter}, {and} \bibinfo{person}{Mona Vij}.} \bibinfo{year}{2017}\natexlab{}.
\newblock \showarticletitle{Graphene-SGX: A Practical Library {OS} for
  Unmodified Applications on {SGX}}. In \bibinfo{booktitle}{{\em 2017 {USENIX}
  Annual Technical Conference ({USENIX} {ATC} 17)}}.
  \bibinfo{publisher}{{USENIX} Association}, \bibinfo{address}{Santa Clara,
  CA}, \bibinfo{pages}{645--658}.
\newblock
\showISBNx{978-1-931971-38-6}
\showURL{%
\url{https://www.usenix.org/conference/atc17/technical-sessions/presentation/tsai}}


\bibitem[\protect\citeauthoryear{Tsai, Son, Jain, McAvey, Popa, and
  Porter}{Tsai et~al\mbox{.}}{2020}]%
        {tsai20civet}
\bibfield{author}{\bibinfo{person}{Chia-Che Tsai}, \bibinfo{person}{Jeongseok
  Son}, \bibinfo{person}{Bhushan Jain}, \bibinfo{person}{John McAvey},
  \bibinfo{person}{Raluca~Ada Popa}, {and} \bibinfo{person}{Donald~E. Porter}.}
  \bibinfo{year}{2020}\natexlab{}.
\newblock \showarticletitle{Civet: An Efficient Java Partitioning Framework for
  Hardware Enclaves}. In \bibinfo{booktitle}{{\em \usenixsec{}}}.
\newblock


\bibitem[\protect\citeauthoryear{Urbanek and Grosjean}{Urbanek and
  Grosjean}{2008}]%
        {rbench}
\bibfield{author}{\bibinfo{person}{Simon Urbanek} {and}
  \bibinfo{person}{Philippe Grosjean}.} \bibinfo{year}{2008}\natexlab{}.
\newblock \bibinfo{title}{R Benchmark}.
\newblock \bibinfo{howpublished}{\url{https://mac.r-project.org/benchmarks/}}.
   (\bibinfo{year}{2008}).
\newblock


\bibitem[\protect\citeauthoryear{Van~Bulck, Piessens, and Strackx}{Van~Bulck
  et~al\mbox{.}}{2017a}]%
        {van17sgx}
\bibfield{author}{\bibinfo{person}{Jo Van~Bulck}, \bibinfo{person}{Frank
  Piessens}, {and} \bibinfo{person}{Raoul Strackx}.}
  \bibinfo{year}{2017}\natexlab{a}.
\newblock \showarticletitle{SGX-Step: A practical attack framework for precise
  enclave execution control}. In \bibinfo{booktitle}{{\em Proceedings of the
  2nd Workshop on System Software for Trusted Execution}}.
  \bibinfo{pages}{1--6}.
\newblock


\bibitem[\protect\citeauthoryear{Van~Bulck, Weichbrodt, Kapitza, Piessens, and
  Strackx}{Van~Bulck et~al\mbox{.}}{2017b}]%
        {van17telling}
\bibfield{author}{\bibinfo{person}{Jo Van~Bulck}, \bibinfo{person}{Nico
  Weichbrodt}, \bibinfo{person}{R{\"u}diger Kapitza}, \bibinfo{person}{Frank
  Piessens}, {and} \bibinfo{person}{Raoul Strackx}.}
  \bibinfo{year}{2017}\natexlab{b}.
\newblock \showarticletitle{Telling your secrets without page faults: Stealthy
  page table-based attacks on enclaved execution}. In \bibinfo{booktitle}{{\em
  26th $\{$USENIX$\}$ Security Symposium ($\{$USENIX$\}$ Security 17)}}.
  \bibinfo{pages}{1041--1056}.
\newblock


\bibitem[\protect\citeauthoryear{Xing, Shanahan, and Leslie-Hurd}{Xing
  et~al\mbox{.}}{2016}]%
        {Xing2016}
\bibfield{author}{\bibinfo{person}{Bin~(Cedric) Xing}, \bibinfo{person}{Mark
  Shanahan}, {and} \bibinfo{person}{Rebekah Leslie-Hurd}.}
  \bibinfo{year}{2016}\natexlab{}.
\newblock \showarticletitle{Intel{\textsuperscript{\textregistered}} Software
  Guard Extensions (Intel{\textsuperscript{\textregistered}} SGX) Software
  Support for Dynamic Memory Management Inside an Enclave}. In
  \bibinfo{booktitle}{{\em Proceedings of the Hardware and Architectural
  Support for Security and Privacy 2016}} {\em (\bibinfo{series}{HASP 2016})}.
  \bibinfo{publisher}{ACM}, \bibinfo{address}{New York, NY, USA}.
\newblock
\showISBNx{978-1-4503-4769-3}


\bibitem[\protect\citeauthoryear{Xu, Cui, and Peinado}{Xu
  et~al\mbox{.}}{2015}]%
        {Xu2015}
\bibfield{author}{\bibinfo{person}{Yuanzhong Xu}, \bibinfo{person}{Weidong
  Cui}, {and} \bibinfo{person}{Marcus Peinado}.}
  \bibinfo{year}{2015}\natexlab{}.
\newblock \showarticletitle{Controlled-Channel Attacks: Deterministic Side
  Channels for Untrusted Operating Systems}. In \bibinfo{booktitle}{{\em
  Proceedings of the 36th {IEEE} Symposium on Security and Privacy (Oakland)}}.
\newblock


\bibitem[\protect\citeauthoryear{{Yahoo}}{{Yahoo}}{2019}]%
        {ycsb}
\bibfield{author}{\bibinfo{person}{{Yahoo}}.} \bibinfo{year}{2019}\natexlab{}.
\newblock \bibinfo{title}{Yahoo! Cloud Serving Benchmark}.
\newblock \bibinfo{howpublished}{\url{https://ycsb.site}}.
  (\bibinfo{year}{2019}).
\newblock


\bibitem[\protect\citeauthoryear{Yu, Shinde, Carlson, and Saxena}{Yu
  et~al\mbox{.}}{2022}]%
        {yu22elasticlave}
\bibfield{author}{\bibinfo{person}{Jason~Zhijingcheng Yu},
  \bibinfo{person}{Shweta Shinde}, \bibinfo{person}{Trevor~E Carlson}, {and}
  \bibinfo{person}{Prateek Saxena}.} \bibinfo{year}{2022}\natexlab{}.
\newblock \showarticletitle{Elasticlave: An efficient memory model for
  enclaves}. In \bibinfo{booktitle}{{\em 31st USENIX Security Symposium (USENIX
  Security 22)}}. \bibinfo{pages}{4111--4128}.
\newblock


\bibitem[\protect\citeauthoryear{ZDNet}{ZDNet}{2020}]%
        {SGXproduct}
\bibfield{author}{\bibinfo{person}{ZDNet}.} \bibinfo{year}{2020}\natexlab{}.
\newblock \bibinfo{title}{Cloud security: Microsoft Azure's SGX VMs hit GA,
  Google's Shielded VM is now default}.
\newblock
  \bibinfo{howpublished}{\url{https://www.zdnet.com/article/cloud-security-microsoft-azures-sgx-vms-hit-ga-googles-shielded-vm-is-now-default/}}.
    (\bibinfo{date}{April} \bibinfo{year}{2020}).
\newblock


\end{thebibliography}
\end{document}